\documentclass[aps,prl,reprint,superscriptaddress]{revtex4-2}
\usepackage[normalem]{ulem}
\usepackage{graphicx}
\usepackage{amsmath,amssymb,mathtools}
\usepackage{amsfonts}
\usepackage{bm}
\usepackage[inline]{enumitem}
\usepackage{xcolor}
\definecolor{apsblue}{RGB}{0,20,150}

\usepackage[
  colorlinks=true,
  linkcolor=apsblue,
  citecolor=apsblue,
  urlcolor=apsblue,
  pdfborder={0 0 0}
]{hyperref}

\graphicspath{{figs/}}

\usepackage{graphicx,epstopdf,amssymb,amsmath,verbatim,hyperref,mathdots,xcolor}
\pdfoutput=1

\newcommand{\vol}{\operatorname{vol}}

\newcommand{\Det}{\operatorname{Det}}

\newcommand{\e}{\mathrm e}

\begin{document}

\author{Sen Mu}
\affiliation{Max Planck Institute for the Physics of Complex Systems, 01187 Dresden, Germany}

\author{Abbas Ali Saberi}
\affiliation{School of Science, Constructor University, Campus Ring 1, 28759 Bremen, Germany}

\author{Martin R. Zirnbauer}
\affiliation{Institute for Theoretical Physics, University of Cologne, Z\"ulpicher Stra{\ss}e 77a, 50937 Cologne, Germany}

\title{Multifractality at the Integer Quantum Hall Transition:\\ Acquittal of the Parabolic Law}

\begin{abstract}
Stationary wave functions at the integer quantum Hall transition are known to be multifractal, but the exact form of the multifractality spectrum has remained a subject of debate. While conformal field theory arguments predict a parabolic law, numerical simulations show deviations from parabolicity. We resolve this discrepancy by pointing out that powers of the local wave intensity fail to obey the Gaussian Free Field and Abelian Fusion Hypothesis assumed in earlier analysis. Rather, due to the non-Abelian nature of the underlying effective field theory, wave intensity correlators are dressed by insertions of background charge distributed uniformly in space. An exact expression for the $q$-moments of point-contact generated eigenstates is presented. Numerical tests are performed for the critical Chalker-Coddington network model on a rectangular torus. Our results are in precise agreement with the predictions of conformal symmetry realized as a level-4 current algebra.
\end{abstract}

\maketitle

\textit{Introduction.}--- Among the quantum phase transitions of Anderson (de-)localization type, the integer quantum Hall (IQH) transition stands out as a benchmark for two-dimensional quantum criticality occurring between distinct phases of disordered topological quantum matter. In a long series of numerical studies \cite{HuckesteinRMP1995,KramerPhysRep2005,SlevinOhtsukiPRB2009,ObuseGruzbergEversPRL2012}, scaling with a divergent localization length has been established for various IQH observables at criticality. In particular, the multifractality of critical wave functions has been found
\cite{EversMirlinRMP2008,EversMildenbergerMirlinPRL2008,ObusePRL2008,SubramaniamGruzbergLudwigPRL2006}
to be a sensitive fingerprint of the IQH transition.

Critical scaling at second-order phase transitions is generally expected to be governed by a conformal field theory (CFT) of some kind. Hence, a major question of condensed matter theory has been: what is the CFT for the IQH transition, and what does it predict for the critical observables? Various proposals have been made over the years \cite{PruiskenNPB1984,Khmelnitskii-1983,PruiskenQHEBook1990,BKSTT00,ReadSaleurNPB2001,LutkenRossPLB2007,IkhlefFendleyCardyPRB2011,BettelheimGruzbergLudwigPRB2012}, the most recent one being that of \cite{ZirnbauerCurrentAlgebra2019}, but a decisive determination is still lacking. On the contrary, the very applicability of CFT principles to Anderson transitions has recently been called into question in a surprising development \cite{PadayasiGruzbergPRL2023}. It is these challenges that we address in the present paper.

Our analytical and numerical probe will be the multifractal nature of critical energy eigenfunctions $\psi(r)$ of the disordered system. Past studies of that diagnostic relied on system-size ($L$) scaling to define the exponents $\Delta_q$ characterizing the multifractality:
\begin{equation}
L^{dq}\,\mathbb{E}\!\left(|\psi(r)|^{2q}\right)\sim L^{-\Delta_q}\qquad (L\to\infty) ,
\label{eq:IPR_standard}
\end{equation}
where $\mathbb{E}(\cdot)$ means taking the average over the disorder. Alas, this definition poses a bottleneck to high-precision numerical studies
\cite{EversMirlinRMP2008,EversMildenbergerMirlinPRL2008,ObusePRL2008,SubramaniamGruzbergLudwigPRL2006}
as there is no direct relation to any CFT observable and the finite-size corrections to the scaling limit (\ref{eq:IPR_standard}) are not analytically known. Nevertheless, based on (\ref{eq:IPR_standard}), researchers have argued \cite{ObusePRL2008,EversMildenbergerMirlinPRL2008} that the $q$-dependence of $\Delta_q$ differs by non-negligible corrections from the parabolic law predicted by CFT; cf.\ (\ref{eq:parabolic}).

In our quest to check CFT predictions (for the IQH multifractality spectrum $\Delta_q$ and other quantities), we are well advised to look for a class of correlation functions where the principles of CFT do apply without any caveat. A good compromise between analytical control and numerical accessibility is to adopt the geometry of a rectangular torus, i.e.\ a strip with finite side lengths, say $L$ and $W$, and periodic boundary conditions in both directions. Given such a torus geometry, can we devise a numerically computable correlator whose infrared limit is analytically accessible as a CFT correlation function?

A viable approach pioneered in \cite{BWZ2014} is to open the isolated IQH system by attaching a point contact with one incoming and one outgoing channel. Then, by constantly forcing probability flux through the incoming channel, one engineers a stationary wave function, $\psi_c \,$, whose disorder-averaged intensity powers $\mathbb{E}(|\psi_c(o)|^{2q})$ depend on the vector $c-o$ from the point $o$ of observation to the position $c$ of the contact. At criticality, the scaling limit of the moments $\mathbb{E}(|\psi_c(o)|^{2q})$ is predicted \cite{BWZ2017} to behave as a genuine CFT two-point function. (To be precise, these moments scale to a two-point function of CFT primary fields for $q = \frac{1}{2} + \mathrm{i} \lambda$, $\lambda \in \mathbb{R}$. The case of real $q \in [0,1]$ is reached by analytic continuation.) Unlike (\ref{eq:IPR_standard}), such a construction makes direct sense in the plane limit, where CFT predicts scaling as a pure power,
\begin{equation}
\lim_{L,W\to\infty} \mathbb{E}(|\psi_c(o)|^{2q}) = \mathrm{const}_{q} \times |c-o|^{-2\Delta_q} ,
\label{eq:IPR_BWZ}
\end{equation}
along with a parabolic law for the scaling dimensions \cite{ZirnbauerCurrentAlgebra2019}:
\begin{equation}
\Delta_q = \frac{1}{n} q(1-q) , \quad n = 4.
\label{eq:parabolic}
\end{equation}

There are several advantages to adopting the plane limit (\ref{eq:IPR_BWZ}) as a CFT-informed (re-)definition of the multifractal scaling dimensions $\Delta_q$. For one, the scaling limit of the defining observable $\mathbb{E}(|\psi_c(o)|^{2q})$ has a closed expression [as a CFT correlator $M_q(c,o)$; see (\ref{eq:torus-M}) below] valid for \emph{any} torus geometry, thereby enabling precise numerical tests free of the uncertainties that plague  (\ref{eq:IPR_standard}). For another, when implemented for the Chalker-Coddington network model \cite{ChalkerCoddington1988} of the IQH transition, the moments $\mathbb{E}(|\psi_c(o)|^{2q})$ enjoy perfect Weyl symmetry ($q \to 1-q$)
\begin{equation}\label{eq:Weyl}
\mathbb{E} \big( |\psi_c(o)|^{2q} \big) =
\mathbb{E} \big( |\psi_c(o)|^{2(1-q)} \big)
\end{equation}
for \emph{any} system geometry and in the \emph{whole} parameter range \cite{BWZ2017} -- a helpful property in monitoring the convergence of the data statistics as the disorder ensemble increases in size. Note also the remarks in \cite{footnote}.

Let us summarize our main findings. (i) Focusing on the critical IQH system, we will present an exact expression for the scaling limit, $M_q(c,o)$, of the wave-intensity moments $\mathbb{E}(|\psi_c(o)|^{2q})$ of point-contact engineered energy eigenstates; that expression separates into a Gaussian free field correlator and a dressing factor. (Tending to a constant in the infinite-volume limit, the latter was not anticipated in earlier work assuming ``Abelian fusion'' \cite{BWZ2017}.) (ii) We test the analytical expression $M_q(c,o)$ by numerical simulations of the Chalker--Coddington network model on a rectangular torus of size $L\times W$. (iii) We show that the approach to the conformal scaling limit $M_q(c,o)$ is anomalously slow: the finite-size corrections die out as $1/\ln(L/L_0)$ for $L,W\to \infty$ (keeping $L/W$ fixed), with $L_0$ of the order of the lattice constant. With these finite-size corrections taken into account, our numerical data are perfectly consistent with the parabolic law (\ref{eq:parabolic}) for the scaling dimensions $\Delta_q$ at the renormalization-group fixed point.

\begin{figure}
    \centering
    \includegraphics[width=1\linewidth]{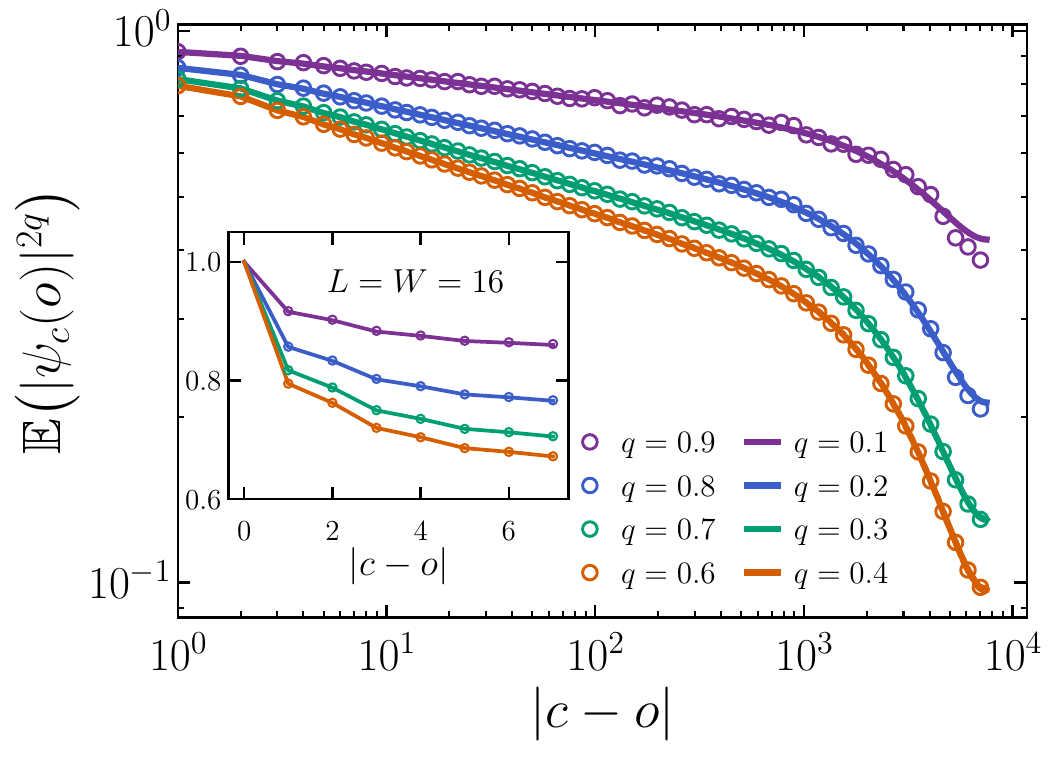}
    \caption{Weyl-symmetry check for the wave-intensity moments $\mathbb{E} \big( |\psi_c(o)|^{2q} \big)$. The disorder average is taken over $N_{\mathrm{dis}} = 10^6$ realizations of the critical CC network model in a torus geometry with $L=15000$ and $W=1500$. The data for $q$ and $1-q$ nearly coincide, indicating statistical convergence. The inset demonstrates that Weyl symmetry holds for all system geometries and sizes, including the smallest ones.}
    \label{fig:weyl_symmetry}
\end{figure}

\textit{Model and observable.}--- Our numerical work is carried out for the Chalker--Coddington (CC) model \cite{ChalkerCoddington1988}. Well established as a description of the single-particle physics of the IQH transition, this model has been amply reviewed, e.g.\ in \cite{KramerPhysRep2005}. Its Hilbert space $\mathbb{C}^{N_{\rm links}}$ is spanned by states $|r\rangle$ on directed links $r$. A single time step is a unitary update $|\Psi_{t+1}\rangle = U\,|\Psi_t\rangle$, where $U$ combines node scattering with link-diagonal random phase factors that model the disorder. The CC model is critical when the probabilities for node scattering are left-right symmetric.

To construct a stationary state of the unitary network dynamics $U$, we employ the scheme of a point contact. Located at some chosen link $c$, the point contact acts as a perpetual source of incoming unit flux and as a sink for outgoing flux (of unit magnitude once the stationary limit has been reached). Let $Q = 1-|c\rangle\langle c|$ denote the projector that implements perfect absorption at the point contact. Then our stationary state $|\psi_c\rangle$
is given by
\begin{equation}
\psi_c(r) \equiv \langle r |\psi_c\rangle = \langle r | (1-QU)^{-1} |c\rangle \quad (r \not= c) .
\end{equation}
By expanding $(1-QU)^{-1}$ as a geometric series, we see that $\psi_c (r)$ is a sum over paths from $c$ to $r$ subject to the constraint of no return to the point $c$ of total absorption. Note that $U |\psi_c\rangle = |\psi_c\rangle$ holds on the orthogonal complement of $|c\rangle$; therefore, our scattering-type state $|\psi_c \rangle$ is stationary with (quasi-)energy $E = 0$.

The observable we study is $\mathbb{E}(|\psi_c(o)|^{2q})$, the disorder-averaged wave-intensity moment of $\psi_c(r)$ taken at an observation point $o$ ($\not = c$). By translation invariance of the disorder distribution, these moments are functions of the difference $c-o$. Due to their meaning as CFT two-point functions \cite{BWZ2017,ZirnbauerCurrentAlgebra2019} in the critical IQH-system, they tend to straight power laws (with scaling exponents $\Delta_q$) in the limit of a planar geometry, $\Sigma = \mathbb{R}^2$. Now, since numerical computations are done in finite size, a much better analytical setting to adopt is the scaling limit for a rectangular torus $\Sigma = \mathrm{S}^1 \times \mathrm{S}^1$ with finite circle lengths $L$ and $W$; that is the setting we will eventually pursue.

As a practical diagnostic of statistical convergence in our numerical simulations of the CC network model, we exploit the exact Weyl symmetry, Eq.~(\ref{eq:Weyl}). In a finite disorder ensemble, any residual mismatch between the data for $q$ and $1-q$ provides a direct estimate of finite-sampling noise and thus helps to identify a reliable data window. We present this test in Fig.~\ref{fig:weyl_symmetry} for a rectangular torus of size $L = 10 W = 15000$: with $N_{\mathrm{dis}}=10^6$ realizations, the symmetry is well satisfied over most of the data range. Deviations are visible for $q$ near 1 and at the largest separations $c-o$; these reflect the expected sensitivity of higher moments to rare events.

\textit{Analytical result: dressed correlator.}--- The conformal scaling limit of $\mathbb{E}(|\psi_c(o)|^{2q})$ results from a modification of the Gaussian Free Field Hypothesis of \cite{BWZ2017} as follows. Let the space manifold $\Sigma$ be a rectangular torus, $\Sigma = \mathrm{S}^1 \times \mathrm{S}^1$ (more generally, $\Sigma$ could be any compact Riemann surface), and let $G(r,r^\prime)$ be the Green's function of the Laplacian $\Delta$ for $\Sigma$:
\begin{equation}
- \Delta G(r,r^\prime) = \delta(r-r^\prime) - \frac{1}{\mathrm{vol}(\Sigma)} \,,
\label{eq:Poisson}
\end{equation}
with $\mathrm{vol}({\Sigma})$ denoting the total area of the surface $\Sigma$. The CFT scaling limit $\mathbb{E}(|\psi_c(o)|^{2q})\to M_q(c,o)$ then is
\begin{equation}
M_q(c,o) = \exp\big( 4\pi G(c,o) \Delta_q \big) S_q(c,o),
\label{eq:torus-M}
\end{equation}
where $S_q(c,o)$, called the dressing factor, is an integral
\begin{equation}
S_q(c,o) = \int_{\Sigma} d^2r \, P_c(r) \, \mathrm{e}^{- (4\pi q/n) \big( G(r,o) - G(r,c) \big) }
\label{eq:dress}
\end{equation}
against the probability measure
\begin{equation}
P_c(r) \, d^2r = C_\Sigma \exp \left( - \frac{4\pi}{n} G(r,c) \right) \frac{d^2 r}{\mathrm{vol}(\Sigma)} , 
\label{eq:prob-meas}
\end{equation}
with constant $C_\Sigma$ chosen so that $\int_\Sigma P_c(r)\, d^2r = 1$.

It must be noted that the solution $G(r,r^\prime)$ to the Poisson equation (\ref{eq:Poisson}) is unique only up to an additive constant. (This just reflects the fact that CFT intrinsically does \emph{not} predict the overall amplitude of lattice-model correlation functions.) The indeterminacy amounts to an unknown ultraviolet scale $a$, which enters as
\begin{equation}\label{eq:UV-limit}
    G(r,r^\prime) = - \frac{1}{2\pi} \ln \frac{\vert r-r^\prime \vert}{a} + \ldots
\end{equation}
in the short-distance limit $|r-r^\prime| \to 0$. We determine $a$ by matching to the Chalker--Coddington numerical data.

Several remarks are needed. (i) The dressing factor $S_q(c,o)$ is easily seen to be invariant under $q \to 1- q$. It follows that our CFT-moments have the required Weyl symmetry $M_q(c,o) = M_{1-q}(c,o)$. (ii) The appearance of $S_q(c,o)$ implies that the naive bosonization rule $|\psi_c(o)|^{2q} \sim \mathrm{e}^{q \varphi(o)}$ with a Gaussian free field $\varphi$ \cite{BWZ2017} misses some effects in finite volume; we are going to show that these are significant.  (iii) We prove in the Supplemental Material~\cite{SuppMat} that formula (\ref{eq:torus-M}) reduces to the power law (\ref{eq:IPR_BWZ}) in the plane limit of $L, W \to \infty$ (keeping $|c-o|$ fixed). To see this informally, note that $S_q(c,o)$ is an integral against the \emph{normalized} measure $P_c(r)\, d^2 r$ and the function $P_c(r)$ \emph{increases} with distance $|r-c|$. Thus, the plane limit pushes the weight of the measure away from $c$ (and $o$) to remote regions of $r$, where the difference $G(r,c) - G(r,o)$ becomes negligible. It follows immediately that $S_q(c,o) \to 1$, and Eq.~(\ref{eq:torus-M}) reduces to the power law (\ref{eq:IPR_BWZ}) for $\Sigma = \mathbb{R}^2$. In contrast, in the scaling limit of a torus, where we keep $|c-o|/L \not= 0$ fixed while sending the system size to infinity, the dressing factor $S_q(c,o)$ remains nontrivial.

As a CFT two-point function of primary fields with scaling dimension $\Delta_q$, the correlator $M_q(c,o)$ takes the form of a scale-invariant function divided by the power $2\Delta_q$ of some characteristic length. Specializing to a torus with periods $L$ and $W$, and keeping the aspect ratio $L/W$ fixed, we may present $M_q(c,o)$ as
\begin{equation}
    M_q(c,o) = L^{-2\Delta_q} \, \Phi_q \left( \frac{c-o}{L} \right) .
\label{eq:collapse}
\end{equation}
Taking $M_q$ from  Eq.\ (\ref{eq:torus-M}), we shall compare the so-defined scaling function $\Phi_q$ with the scaling limit (i.e.\ $L \to \infty$ and $|c-o| \to \infty$, with $|c-o|/L$ fixed) of $\mathbb{E}(|\psi_c(o)|^{2q})$.

\begin{figure}
    \centering
    \includegraphics[width=\columnwidth]{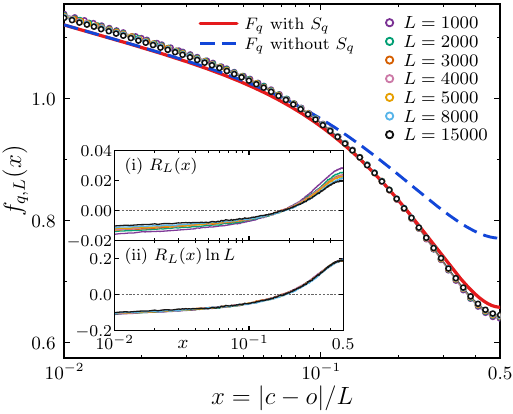}
    \caption{Comparison between the scaled numerical data $f_{q,L}(x = |c-o|/L) \equiv L^{2\Delta_q}\mathbb{E}\left(|\psi_c(o)|^{2q}\right)$ and the predicted scaling function $F_q(x)$ for $q = 0.1$, $L/W=10$ and varying values of $L$. All numerical data are rescaled using the \emph{same} CFT exponent, $\Delta_q=q(1-q)/4$, independent of system size. The analytical curves use the fixed UV scale $a=0.13$, the red solid curve is the full prediction in Eq.~(\ref{eq:torus-M}) including the dressing factor $S_q$, while the blue dashed curve omits $S_q$. Inset (i) displays the logarithmic residual $R_L(x)\equiv\ln F_q(x)-\ln f_{q,L}(x)$. Inset (ii) shows $R_L(x)\ln L$ over the same $x$-range. The collapse for different $L$ is consistent with a leading finite-size correction of the form $R_L(x)=\rho(x)/\ln L$.}
    \label{fig:2}
\end{figure}

\textit{Numerical results.}--- We begin by showing that the factor $S_q$, though a finite-volume effect that disappears for $L \to \infty$ when $|c-o|$ is kept fixed, cannot be neglected in general. To demonstrate its quantitative role, we consider a large aspect ratio $L/W = 10$ and set $q = 0.1$ for concreteness. For simplicity and without much loss, we take the vector $c-o$ to be parallel to the long axis (of length $L$); the scaling function $\Phi_q$ in (\ref{eq:collapse}) then becomes a function $F_q(|c-o|/L)$ of the distance $|c-o|$. Fig.\ \ref{fig:2} shows the full scaling function $F_q$ (red solid curve) and the same function with the scale-invariant dressing factor $S_q$ omitted (blue dashed curve); the details of the numerical implementation of $S_q$ are given in the Supplemental Material~\cite{SuppMat}. Clearly, there is a large difference between the two curves, and the latter of the two does not approximate the function which is obtained by collapsing the numerical data (see below). That conclusion remains true for other values of $L/W$ and $q$.

We turn to the discussion of the data that we have obtained for $\mathbb{E}(|\psi_c(o)|^{2q})$ by numerical simulation. Our primary goal here is to compare $F_q$ with the values computed numerically for varying system sizes. To remove the characteristic scale dependence, we rescale the data using for every system size the \emph{same} CFT exponent $\Delta_q\,$: 
\begin{equation}
    f_{q,L}\left(\frac{|c-o|}{L}\right) \equiv L^{2\Delta_q} \mathbb{E}(|\psi_c(o)|^{2q}). 
\end{equation}
Thus, any remaining size dependence is exposed as a finite-size deviation from the scaling function $F_q$ predicted to emerge at the renormalization-group fixed point of the critical system. The resulting observable $f_{q,L}$ is shown in Fig.~\ref{fig:2} for system sizes up to $L=15000$. A notable feature (best visible in the inset) is that all curves $f_{q,L}(x)$ cross at a point $x_c \approx 0.18$ nearly independent of size $L$! Now on basic grounds, we must allow for the CFT scaling function $F_q(x)$ to differ from the Chalker-Coddington lattice data $f_{q,L}(x)$ (non-universal at short distances) by an $x$-independent multiplicative constant. Since the overall normalization of $F_q$ is controlled by the unknown ultraviolet scale $a$ in (\ref{eq:UV-limit}), we fix $a$ by matching $F_q(x_c)$ to $f_{q,L}(x_c)$. This gives $a \approx 0.13$. Our finding then is that the numerical curves $f_{q,L}(x)$ move systematically toward the CFT prediction $F_q(x)$ as $L$ increases, although the approach is anomalously slow; see also Ref.~\cite{DresselhausSbierskiGruzbergAOP2021}. We quantify this behavior by considering the logarithmic residual
\begin{equation}
    R_L(x) \equiv \ln F_q(x)-\ln f_{q,L}(x).
\end{equation}
Inset (i) of Fig.~\ref{fig:2} displays $R_L(x)$ over the full range of scaled separations $x$. [By construction, $R_L(x_c) = 0$.] The magnitude of $R_L(x)$ decreases systematically with increasing $L$, while its sign changes at $x_c$. Inset (ii) shows that multiplication of $R_L(x)$ by $\ln L$ collapses all curves onto a single one. This collapse strongly suggests that the leading finite-size correction to the scaling limit is of the form $R_L(x) = {\rho(x)}/{\ln L}$ with size-independent $\rho$. Thus, our numerical data show logarithmic convergence to the CFT scaling function $F_q(x)$ at fixed $x$ as $L\to\infty$. Analyzed in this way, the data vindicate the parabolic law (\ref{eq:parabolic}) in the conformal scaling limit.

Eq.~(\ref{eq:torus-M}) also shows that omitting $S_q$ in fits of  finite-volume data can generate \emph{apparent} non-parabolicity, even though the properly defined scaling dimensions are exactly parabolic in $q$. Since $\ln S_q(c,o)$ enters additively in $\ln M_q(c,o)$, an undressed ansatz linear in $G(c,o)$ absorbs its distance-dependent part into the fitted slope. For $|c-o|$ comparable to the system size, $\ln S_q(c,o)$ is nonlinear in $q(1-q)$ and will produce a non-parabolic contribution to the \emph{effective} exponent. This bias is quantified in the Supplemental Material~\cite{SuppMat}.

As a related caveat, the finite-size corrections to the scaling limit $F_q$ are \underline{not} eliminated by restricting the fit to an $x$-region where $S_q$ is negligible. In other words, the percent-level deviations reported in \cite{BWZ2014} must not be interpreted as a failure of the CFT prediction (\ref{eq:IPR_BWZ}, \ref{eq:parabolic}).

\textit{Sketch of proof.}--- While a first-principles derivation of (\ref{eq:torus-M}) is beyond the scope of this Letter, we shall now highlight the key points (relegating the details to the Supplemental Material~\cite{SuppMat}). The central object is a Gaussian free field (GFF), $\varphi$, which enters via the bosonization rule $| \psi_c |^2 \sim \mathrm{e}^{\varphi}$. The zero-mode integral for $\varphi$ extracts from the field-theory image of the point-contact projector $Q$ a factor $\mathrm{e}^{-q \varphi(c)}$ balancing the observable $|\psi_c(o)|^{2q} \sim \mathrm{e}^{q \varphi(o)}$. Crucially, it also produces a factor $\xi(c)\, \eta(c)$ of fermions from the parent field theory, the WZW model of \cite{ZirnbauerCurrentAlgebra2019}. Integration over the fields $\xi, \eta$ yields the determinant of a Witten-deformed Laplacian (of Dolbeault type), $D_\varphi^\dagger D_\varphi^{\vphantom{\dagger}}$ with $D_\varphi = \mathrm{e}^{\varphi/2} \partial_{\bar z} \,\mathrm{e}^{-\varphi/2}$,  $D_\varphi^\dagger = - \mathrm{e}^{-\varphi/2} \partial_z \,\mathrm{e}^{\varphi/2}$, and complex coordinates $z, \bar{z}$ for the torus $\Sigma$. The determinant is taken over the Hilbert space of square-integrable functions $f$ on $\Sigma$, with Dirichlet condition $f(c) = 0$ for absorption at the contact $c$. Correcting the naive two-point GFF correlator $\langle \mathrm{e}^{q \varphi(o)} \mathrm{e}^{-q \varphi(c)} \rangle$, this determinant contributes a factor $\mathrm{e}^{\varphi(c)} \int_\Sigma d^2 r\, \mathrm{e}^{-\varphi(r)} / \mathrm{vol}(\Sigma)$, where $\mathrm{e}^{\varphi(c)}$ reflects  the Dirichlet condition at $c$, while the factor $\mathrm{e}^{-\varphi(r)}$ with variable position $r$ originates from a ``virtual'' zero mode (namely, $D_\varphi^\dagger \, \mathrm{e}^{-\varphi/2} = 0$).

Summarizing the situation from the reduced perspective of the Gaussian free field $\varphi$, there is a charge $q = \frac{1}{2} + \mathrm{i}\lambda$ at the observation point $o$, a charge $1-q = \frac{1}{2} - \mathrm{i}\lambda$ at the contact point $c$, and a charge $-1$ at the variable position $r$. Given this configuration with overall charge neutrality, one takes a Gaussian free field correlator,
\begin{equation}\label{eq:GFF-3}
M_q(c,o) = \int_\Sigma \frac{d^2 r}{\mathrm{vol}(\Sigma)} \left\langle \mathrm{e}^{q \varphi(o)} \mathrm{e}^{(1-q)\varphi(c)} \mathrm{e}^{-\varphi(r)} \right\rangle ,
\end{equation}
which becomes the expression (\ref{eq:torus-M}) when the field stiffness is taken from the WZW current algebra of level $n = 4$. 

Let us emphasize that Eq.~(\ref{eq:GFF-3}) is meant for the \emph{flat} torus considered in our numerical work. On a closed surface $\Sigma$  with curvature form $R$, the GFF action functional would also contain a Polyakov background charge term proportional to $\int_\Sigma \varphi R$.

\textit{Discussion and conclusion.}---
To arrive at a full understanding of the critical behavior at the IQH transition, two questions must be answered: (i) what is the exact nature of the stable flow \emph{into} the fixed point of the renormalization group (RG), and (ii) how does the localization length diverge due to the unstable flow \emph{out of} it? Clearly, one cannot give a quantitative answer to question (ii) without having answered (i) first. Therefore, our focus here was on (i), leaving (ii) for future work.

In a follow-up article, we will present numerical data for the critical CC network model with random node-scattering probabilities (a.k.a.\ ``geometric disorder''). These numerical results exhibit the \emph{same} infrared behavior, namely $1/\ln L$ creep toward the universal RG-fixed point with current-algebra level $n = 4$. The difference is that the geometrically disordered model sits further away from the fixed point; therefore, limited by numerically tractable system sizes, one finds it even more difficult to get close to it. In view of this difficulty, we dispute the claim \cite{KNS19} that universality breaks down by geometric disorder inducing a \emph{line} of RG-fixed points.

To reinforce our message, let us comment once more on (\ref{eq:IPR_standard}). That scaling ansatz is suboptimal in two respects: (i) Packaging all dependence into a single length scale $L$ (when the typical system geometries used in numerical simulations involve more than one scale), it does not take into account boundary and geometry effects, which we have seen to be non-negligible in general. (ii) The ansatz (\ref{eq:IPR_standard}) does not allow for the complication that the RG flow into the fixed point of the IQH transition is anomalously slow, making for finite-$L$ corrections that creep toward zero like $1/ \ln L$ (and, in all likelihood, are non-parabolic in $q$). So, lacking an accurate finite-size scaling theory, researchers have been (mis-)led to conclude that the $q$-dependence of $\Delta_q$ is non-parabolic and, furthermore, that there is even a violation of conformal symmetry.

Our formula (\ref{eq:torus-M}) takes care of the first issue but, unfortunately, not of the second one. Adopting the Pruisken-Khmelnitskii (PK) scaling picture \cite{PruiskenNPB1984,Khmelnitskii-1983}, which assumes one irrelevant coupling $\sigma_{xx}(L) - \sigma_{xx}^\ast \,$, we have attempted (devoid of direct theoretical justification) to fit our numerical data by making in Eq.~(\ref{eq:torus-M}) the simple replacement $4 = n \equiv 2\pi\sigma_{xx}^\ast \to 2\pi\sigma_{xx}(L)$ with a running longitudinal mean conductivity $\sigma_{xx}(L)$. Displayed in the Supplemental Material~\cite{SuppMat}, the resulting fits turn out to be unsatisfactory on two accounts. For one, the fitted value $\sigma_{xx}(L)$ has a weak dependence on the data range used in the fit, leaving a systematic uncertainty at the level of 2 percent. For another, the offset $\delta \equiv \sigma_{xx}(L) - \sigma_{xx}^\ast$ decreases as $1 / \ln L$. Assuming one-parameter scaling for the stable RG flow (as is done in the PK scaling picture), one runs into a contradiction: $1/\ln L$ behavior is incompatible with the analyticity of a beta function $\beta(\sigma_{xx})$ which, to leading order, must be odd in $\delta$. We take this as strong evidence that the RG flow into the IQH fixed point cannot be understood quantitatively by simply renormalizing a single coupling $\sigma_{xx}$.

The main lesson we have learned is that the Gaussian Free Field and Abelian Fusion Hypothesis of \cite{BWZ2017} has to be modified: what is a true two-point function $M_q(c,o)$ in the parent-CFT turns into an (integrated) three-point function in the reduced theory with only a Gaussian free field. Put differently, even though the vertex fields $|\psi(r)|^{2q} \sim \mathrm{e}^{q \varphi(r)}$ do combine with each other by Abelian fusion, any charge-neutral correlation function of such fields gets ``dressed'' by the coupling to other sectors of the non-Abelian parent theory. In hindsight, this should have been expected: the phenomenon of $m$-point functions being dressed up to $(m+p)$-point functions ($p > 0$) is well known from the correspondence \cite{RT05,HS07,GKR25} between the $\mathrm{H}^3$-WZW model and Liouville theory.

Thus, we face the complication that valid correlation functions of ``Abelian-fusion observables'' are dressed by operator insertions. This is sure to be a general feature that likewise affects other 2D Anderson-critical points, in particular that of the class-$C$ model (namely the CC network model with two channels per link and $\mathrm{SU}(2)$-Haar disorder). Moreover, we expect these complications to become even more pressing for the case of ``generalized multifractality observables'' \cite{KCGM21}, where further Abelian-fusion fields and their coupling to other sectors come into play. These issues and their resolution may well explain the CFT violations argued, e.g., in \cite {PadayasiGruzbergPRL2023}. Our bottom line then is that we plead not only for acquittal of the parabolic law, but also for acquittal of the principle of conformal symmetry at Anderson transitions.

\textit{Acknowledgments.} S.M. thanks J.\ Karcher for helpful discussions on numerical simulations. This work was funded by the German Research Foundation (DFG) under Projects No.~557852701 (A.A.S.) and ZI 513/3-1 (M.R.Z.). It was also supported by the Advanced Study Group ``Strongly Correlated Extreme Fluctuations'' at the Max Planck Institute for the Physics of Complex Systems, Dresden (2024/25) \cite{pks_asg2024}.


\clearpage
\onecolumngrid
\clearpage
\hypersetup{pageanchor=false}
\setcounter{page}{1}
\setcounter{section}{0}
\setcounter{subsection}{0}
\setcounter{subsubsection}{0}
\setcounter{equation}{0}
\setcounter{figure}{0}
\setcounter{table}{0}
\setcounter{secnumdepth}{3}
\renewcommand{\theequation}{S\arabic{equation}}
\renewcommand{\thefigure}{S\arabic{figure}}
\renewcommand{\theHequation}{SM.eq.\arabic{equation}}
\renewcommand{\theHfigure}{SM.fig.\arabic{figure}}
\renewcommand{\theHsection}{SM.sec.\arabic{section}}
\renewcommand{\theHsubsection}{SM.subsec.\arabic{section}.\arabic{subsection}}

\begin{center}
{\large\bfseries Supplementary Material for\\[0.5ex]
\textit{``Multifractality at the Integer Quantum Hall Transition:\\
Acquittal of the Parabolic Law''}\par}
\vspace{0.85em}
Sen Mu$^{1}$, Abbas Ali Saberi$^{2}$, and Martin R.\ Zirnbauer$^{3}$\\[0.65em]
{\small
$^{1}$Max Planck Institute for the Physics of Complex Systems,\\
01187 Dresden, Germany\\
$^{2}$School of Science, Constructor University,\\
Campus Ring 1, 28759 Bremen, Germany\\
$^{3}$Institute for Theoretical Physics, University of Cologne,\\
Z\"ulpicher Stra{\ss}e 77a, 50937 Cologne, Germany\\[0.6em]
(\today)\par}
\end{center}

\begingroup
\makeatletter
\let\SMorigcontentsline\contentsline
\newif\ifSMseenrefs
\SMseenrefsfalse
\def\SMreftitle{\numberline {}References}
\renewcommand{\contentsline}[4]{%
  \def\SMthistitle{#2}%
  \ifx\SMthistitle\SMreftitle
    \ifSMseenrefs
      \SMorigcontentsline{#1}{#2}{#3}{#4}%
    \else
      \SMseenrefstrue
    \fi
  \else
    \SMorigcontentsline{#1}{#2}{#3}{#4}%
  \fi
}
\tableofcontents
\makeatother
\endgroup

\makeatletter
\let\SMoriglabel\label
\renewcommand{\label}[1]{%
  \def\SMtarget{#1}\def\SMlastbib{LastBibItem}%
  \ifx\SMtarget\SMlastbib \SMoriglabel{SM:LastBibItem}%
  \else \SMoriglabel{#1}\fi
}
\makeatother

\section{Observable, notation, short-distance limit}

Building on the proposal of \cite{SM:Zi-19,SM:Zi-21}, this Supplementary Material details the analysis underlying our prediction, $M_q(c,o)$, for the stationary-wave intensity moments $\mathbb{E}( |\psi_c(o)|^{2q})$ of the Chalker-Coddington network model at criticality, assuming the conformal scaling limit for a flat rectangular torus with point contact. We describe the numerical implementation and present the lengthy analytical arguments leading to the formula for $M_q(c,o)$. Special attention is paid to the computation of a fermion determinant with ``virtual'' zero mode. We show how the formula for the finite torus goes over to the known formulas for the infinite cylinder and the infinite plane in the corresponding limits. We also describe an (unsuccessful) attempt to fit the finite-size numerical data with an ansatz motivated by the Pruisken-Khmelnitskii scaling picture.

We work on the flat rectangular torus $\Sigma=\mathbb R^2/(L\mathbb Z\oplus W\mathbb Z)$, of area $\vol(\Sigma)=LW$. The point contact is located at position $c$, the observation point at position $o$ ($\ne c$). If $Q=1-|c\rangle\langle c|$, and $U$ is the evolution operator for one time step, the stationary state generated by one unit of flux (permanently injected into the network through $c$) is
\begin{equation}
    |\psi_c\rangle=(1-QU)^{-1}|c\rangle .
\end{equation}
Expanded as a geometric series,
\begin{equation}
    (1-QU)^{-1} = \sum_{k \geq 0} (QU)^k ,
\end{equation}
the resolvent operator $(1-QU)^{-1}$ has an interpretation as a sum over paths. The projector $Q$ deletes from the path sum for $|\psi_c \rangle$ all paths that return to the contact $c$ (i.e., the contact is a perfectly absorbing sink). 

Our observable of interest is the disorder average $\mathbb{E}\big( |\psi_c(o)|^{2q} \big)$ for $\psi_c(o) \equiv \langle o | \psi_c \rangle$. Its critical scaling limit is denoted by $M_q(c,o)$. Section~\ref{SM:app:numerical_implementation} uses the symbol $M_q^{\rm latt}$ for the approximation of $M_q$ by discretization on a lattice. From the perspective of (unitary) representation theory, the ``good'' $q$-values are $q = \frac{1}{2} + \mathrm{i} \lambda$ (for $\lambda \in \mathbb{R}$). In our numerical work, we consider real values of $q$ in the interval $0\leq q\leq1$ (by analytic continuation from the axis $\mathrm{Re}\, q = \frac{1}{2}$). 

We now state our prediction for $M_q(c,o)$ from conformal field theory (CFT) on the flat rectangular torus $\Sigma$. The Green's function $G(r,r^\prime)$ of the Laplacian $\nabla^2 \equiv \mathrm{div} \circ \mathrm{grad}$ for $\Sigma$ is defined (up to an additive constant) by 
\begin{equation} \label{SM:eq:Poisson}
    -\nabla^2 G(r,r')=\delta(r-r')-\frac{1}{\mathrm{vol}(\Sigma)} .
\end{equation}
[The Laplacian $\nabla^2$ has a one-dimensional co-kernel given by the constant (density) functions; the term $1/\mathrm{vol}(\Sigma)$ is subtracted to put the right-hand side into the orthogonal complement of that co-kernel.] Our prediction for the scaling limit (where all lengths are sent to infinity while length ratios are kept fixed) of $\mathbb{E}(|\psi_c(o)|^{2q})$ is 
\begin{equation}\label{SM:eq:torus-M}
    M_q(c,o) =  \big( \mathrm{e}^{2\pi G(c,o)} \big)^{2\Delta_q} C_\Sigma \int_\Sigma \!\! d^2r \; \exp\!\left( -\frac{4\pi}{n} q \, G(r,o) - \frac{4\pi}{n} (1-q) \, G(r,c) \right),
\end{equation}
with scaling dimensions 
\begin{equation}\label{SM:eq:parabolic}
 \Delta_q = \frac{q(1-q)}{n}, \qquad n = 4,
\end{equation}
and normalization constant $C_\Sigma$ determined by
\begin{equation}\label{SM:eq:normalization}
    1 = M_{q=0} (c,o) = C_\Sigma \int_\Sigma d^2r \, \exp\!\left( - \frac{4\pi}{n} G(r,c) \right).
\end{equation}
Referred to as the ``dressing factor'', the integral in Eq.~(\ref{SM:eq:torus-M}) is non-trivial in finite geometry [$\mathrm{vol}(\Sigma) < \infty$], where it corrects the naive two-point function $( \mathrm{e}^{2\pi G(c,o)})^{2\Delta_q}$ for a Gaussian free field while preserving the symmetry under the Weyl group transformation $q \to 1-q$, which is known \cite{SM:BWZ-17} to be an exact symmetry of $\mathbb{E} ( |\psi_c(o)|^{2q} )$.  

The integral in Eq.~(\ref{SM:eq:normalization}) is independent of the location $c$ of the point contact by $G(r,c) = g(r-c)$ due to the translation symmetry of the flat torus. In the limit of infinite volume, the integral goes to infinity (while $C_\Sigma$ goes to zero to keep the product fixed at value 1). As a CFT two-point function of primary fields with scaling dimension $\Delta_q\,$, the scaling limit $M_q(c,o)$ behaves at short distances as 
\begin{equation}
    M_q(c,o) \stackrel{o\to c}{\longrightarrow} (|c-o|/a)^{-2\Delta_q} ,
\end{equation}
where the ultraviolet scale $a$ is determined by the condition $\int_\Sigma d^2 r \ G(r,r^\prime) = 0$ for the covariance $\langle \varphi(r) \varphi(r^\prime) \rangle$ of a Gaussian free field $\varphi(r)$ with zero mode removed (see Section \ref{SM:app:justification}). In contradistinction, the observable $\mathbb{E}( |\psi_c(o)|^{2q} )$ has the short-distance limit
\begin{equation}\label{SM:eq:UV-limit}
    \mathbb{E}\big( |\psi_c(o)|^{2q} \big) \stackrel{o\to c}{\longrightarrow} 1 .
\end{equation}
The discrepancy for $|c-o| \to 0$ is inevitable, as conformal field theory cannot be expected to account for non-universal lattice effects at short distances, which get eliminated by taking the scaling limit. [The limit (\ref{SM:eq:UV-limit}) is in fact non-universal and changes, e.g., when the absorption strength of the contact is modified.] This means that when comparing the numerical data for $\mathbb{E}( |\psi_c(o)|^{2q})$ to the predicted scaling limit $M_q(c,o)$, we must allow for adjustment by an overall multiplicative constant. 

The following sections contain material in support of the analytical prediction (\ref{SM:eq:torus-M}) and its numerical implementation. Section~\ref{SM:app:numerical_implementation} describes how we calculate  $M_q(c,o)$ numerically and how we fix an unknown ultraviolet scale by comparison with Chalker--Coddington data. Section~\ref{SM:app:limits} shows that the known formulas for $M_q(c,o)$ on the cylinder $\Sigma = \mathbb{R} \times \mathrm{S}^1$ and the plane $\Sigma = \mathbb{R}^2$ are recovered from Eq.~(\ref{SM:eq:torus-M}) by taking $L \to \infty$ and $L,W \to \infty$, respectively. Section~\ref{SM:app:justification} gives a (partly heuristic) derivation of formula (\ref{SM:eq:torus-M}).  Section~\ref{SM:app:fake_nonparabolicity} demonstrates that omission of the dressing factor $S_q$ in finite-geometry fits produces an apparent non-parabolic contribution to the exponents $\Delta_q$ that are obtained by fitting. Based on a heuristic attempt to describe the finite-size approach to the scaling limit $M_q(c,o)$ by a one-parameter flow [replacing $1/n$ by $X(L)$], Section~\ref{SM:app:fits} compares three finite-size fitting procedures and shows that the full-range fit including the dressing factor extrapolates to a limit $X^\ast = X(L\to\infty)$ consistent with $1/4$. This attempt lacks theoretical justification but is included here for completeness.
\bigskip
\section{Evaluation of torus formula by discretization}\label{SM:app:numerical_implementation}

Our numerical simulations are carried out for the critical Chalker-Coddington (CC) network model on a flat rectangular torus of size $L \times W$. We now describe the procedure by which we compare the numerical data for $\mathbb{E}( |\psi_c(o)|^{2q})$ to the analytical prediction $M_q(c,o)$ in Eq.\ (\ref{SM:eq:torus-M}).

Recall that our stationary states $\psi_c$ stem from a point contact at location $c$ and are observed at position $o$. For simplicity and without loss, we place both $c$ and $o$ on the same axis in the long torus direction (of length $L$).

The main ingredient in the predicted formula (\ref{SM:eq:torus-M}) is the Green's function $G(r,r^\prime)$ defined (up to an arbitrary additive constant) by Eq.\ (\ref{SM:eq:Poisson}). By the translation invariance of the flat torus, $G(r, r^\prime)$ depends only on the difference vector $v = r-r^\prime$, so we write $G(r,r^\prime) \equiv g(r-r^\prime) = g(v)$. Using the known properties of Jacobi theta functions, one finds that the continuum Green's function is given by
\begin{equation}\label{SM:eq:Jacobi}
    \chi(x,y) = - \frac{1}{2\pi} \ln \bigg\vert \sum_{n \in \mathbb{Z} + \frac{1}{2}} (-1)^{n-\frac{1}{2}} \exp \left( - \frac{\pi}{LW}(x - nL)^2 + 2\pi\mathrm{i} n \frac{y}{W} \right) \bigg\vert + C_{L,W,a} 
\end{equation}
in (local) Cartesian coordinates $-L/2 \leq x \leq L/2$  and $-W/2 \leq y \leq W/2$ for the torus. In order for the short-distance limit to be 
\begin{equation}
    \chi(x,y) \stackrel{x,y\to 0}{\longrightarrow} - \frac{1}{2\pi} \ln \big( \sqrt{x^2 + y^2} / a \big) ,
\end{equation}
we make the choice of additive constant
\begin{equation}\label{SM:eq:CLWa}
    C_{L,W,a} = \frac{1}{2\pi} \ln \bigg( \frac{4\pi a}{W} \sum_{n \in \mathbb{N} + \frac{1}{2}} (-1)^{n-\frac{1}{2}} n \, \mathrm{e}^{-\pi n^2 L /W}  \bigg) .
\end{equation}
We note that $\chi(x,y)$ is invariant under scale transformations $x \to sx$, $y\to sy$, $L \to sL$, $W\to sW$, $a \to sa$. 

Now in principle, we could compute $g(v) = \chi(x(v),y(v))$ by carrying out the Gauss sum in the expression (\ref{SM:eq:Jacobi}). However, since we are given the numerical data in \emph{discrete} form by the CC network model, and since the integral in (\ref{SM:eq:torus-M}) is not easy to do analytically, we choose a different approach: we solve the Poisson equation (\ref{SM:eq:Poisson}) by discretization on a lattice with nearest-neighbor Laplacian $\nabla^2$. To facilitate the comparison with the numerical data, we choose a square lattice with the same number of lattice sites as the CC model on the $L \times W$ torus. The additive constant in the Green's function is determined by imposing the condition $\sum_v g(v) = 0$ [at a later stage, we shift $g(v) \to g(v) + \mathrm{const}$ to implement the proper short-distance limit; see the discussion above]. Changing the integration variable $r$ in (\ref{SM:eq:torus-M})  to $v \equiv r-c$, we have
\begin{equation}
G(r,o) = g(v+c-o) , \quad G(r,c) = g(v) .
\end{equation}
The discrete approximation to our formula (\ref{SM:eq:torus-M}) now reads
\begin{equation}
M_q^{\rm latt}(c,o) = \mathrm{e}^{{\pi q(1-q)} g(c-o)} S_q^{\rm latt}(c,o) ,
\label{SM:eq:lattice_moment_SM}
\end{equation}
where we have put $4\pi / n = \pi$. To compute the dressing factor $S_q\,$, we replace the continuum integral over the dipole vector $v = r - c$ by a lattice sum over the points of the discrete torus:
\begin{equation}
S_q^{\rm latt}(c,o) = \mathcal{N}^{-1} \sum_v \mathrm{e}^{-\pi q g(v+c-o) -\pi (1-q) g(v)} ,
\label{SM:eq:screening_sum_SM1}
\end{equation}
normalized by $\mathcal{N} = \sum_v \mathrm{e}^{- \pi g(v)}$ so that $S_{q=0}^{\rm latt}(c,o) = 1$.

The remaining non-universal parameter in Eq.~\eqref{SM:eq:lattice_moment_SM} is the short-distance scale $a$ that enters as an additive constant in the Green's function. We adjust $a$ by matching the analytical prediction to the nearly size-independent crossing point, $x_c\approx0.18$, of the rescaled numerical curves $L^{2\Delta_q}\mathbb{E}(|\psi_c(o)|^{2q})$, plotted as functions of $|c-o|/L$ for different $L$. This yields $a\approx0.13$. (Alternatively, and equivalently, we could multiply the rescaled numerical curves by a constant multiplicative factor to align the numerical data with the universal scaling limit predicted by CFT.) 

\bigskip
\section{From the torus to the infinite cylinder and plane}\label{SM:app:limits}

Here we show that the formula (\ref{SM:eq:torus-M}) for $M_q(c,o)$ contains the known formula \cite{SM:BWZ-14} for the infinite cylinder ($\Sigma = \mathbb{R} \times \mathrm{S}^1$) as a limit. For that purpose, we re-organize the expression for $M_q(c,o)$ as
\begin{equation}
M_q(c,o) = \exp\big( 4\pi G(c,o) \Delta_q \big) \, S_q(c,o), 
\end{equation}
with dressing factor
\begin{equation}
    S_q(c,o)= \frac{\int_\Sigma d^2 r\, H(r,o)^q H(r,c)^{1-q}} {\int_\Sigma d^2 r \,H(r,c)},    \quad H(r,\bullet) = \exp \left( - \frac{4\pi}{n} G(r,\bullet) \right) .
\end{equation}
We take the limit of $L \to \infty$, keeping $c-o$ and $W$ fixed. In that limit, the Gauss sum (\ref{SM:eq:Jacobi}) for the Green's function reduces to a sum of two terms ($n = \pm 1/2$), which combine to a hyperbolic sine function:
\begin{equation}
    \chi(x,y) \stackrel{L\to\infty}{\longrightarrow} - \frac{1}{2\pi} \ln \left\vert \frac{W}{\pi a} \, \sinh  \big( \frac{\pi z}{W} \big) \right\vert \quad (z = x + \mathrm{i}y) .
\end{equation}
Thus, with the identification $z \equiv x(c-o) + \mathrm{i} y(c-o)$, the exponential factor in $M_q(c,o)$ converges to
\begin{equation}
    \exp\big( 4\pi G(c,o) \Delta_q \big) \stackrel{L\to\infty}{\longrightarrow} \left\vert \frac{W}{\pi a} \, \sinh  \big( \frac{\pi z}{W} \big) \right\vert^{-2\Delta_q} ,
\end{equation}
which is the formula proposed for $M_q(c,o)$ in \cite{SM:BWZ-14}.

It remains to show that the dressing factor $S_q(c,o)$ converges to unity in the same limit. Intuitively, this comes about because the main weight of the distribution  $H(r,c)\, d^2 r$ is concentrated at positions $r$ far from $c$, where the distance between $c$ and $o$ becomes negligible, with the consequence that $H(r,o)^q H(r,c)^{-q} \approx 1$. 

This intuition can be made rigorous by applying Lebesgue's Dominated Convergence Theorem, as follows. Fix a standard torus $\Sigma_0$ of size $L_0 \times W$, and let $\Psi_L : \; \Sigma_0 \to \Sigma$ be a mapping that stretches the long torus cycle by the scale factor $L/L_0$. Let $o = \Psi_L(o^\prime)$  and $c = \Psi_L(c^\prime)$. Then, by pullback from $\Sigma$ to $\Sigma_0$ we have
\begin{equation}
 S_q(c,o)= \frac{\int_{\Sigma_0} d^2 r^\prime \, (\Psi_L^\ast H)(r^\prime,o^\prime)^q (\Psi_L^\ast H)(r^\prime,c^\prime)^{1-q}}
 {\int_{\Sigma_0} d^2 r^\prime (\Psi_L^\ast H)(r^\prime,c^\prime)} .
\end{equation}
To retain control on $\Psi_L^\ast H \equiv H \circ \Psi_L$ with increasing $L$, we multiply both numerator and denominator by the same scale factor $\mathrm{e}^{(4\pi / n) \, C_{L,W,a}}$ with the constant $C_{L,W,a}$ defined in Eq.~(\ref{SM:eq:CLWa}). For $0 \leq \mathrm{Re}(q) \leq 1$ the integrand in the numerator is then bounded in absolute value by an $L^1$-integrable function on $\Sigma_0$. In the limit of $L \to \infty$, the numerator integrand converges pointwise to the denominator integrand. Dominated convergence lets us take the limit $L \to \infty$ inside the integrands, which yields the desired result $S_q(c,o) \to 1$. 

Thus for the infinite cylinder $\Sigma = \mathbb{R} \times \mathrm{S}^1$ with circumference $W$ we have
\begin{equation}
    M_q(c,o) \stackrel{L\to\infty}{\longrightarrow} \left\vert \frac{W}{\pi a} \, \sinh  \big( \frac{\pi z}{W} \big) \right\vert^{-2\Delta_q} , \quad z \equiv x(c-o) + \mathrm{i} y(c-o) .
\end{equation}
As a corollary, if we now take $W \to \infty$, we obtain the result
\begin{equation}
    M_q(c,o) \stackrel{L,W\to\infty}{\longrightarrow} \big( \vert c-o \vert / a \big)^{-2\Delta_q}
\end{equation}
for the limit of the Euclidean plane, $\Sigma = \mathbb{R}^2$.

\bigskip
\section{Derivation of the Formula for the Scaling Function}\label{SM:app:justification}

\subsection{Brief review of earlier work}

Our first task is to put the present work in perspective with the earlier work of \cite{SM:Zi-19}. We must begin with a disclaimer: to derive the scaling limit (\ref{SM:eq:torus-M}) from the conformal field theory of the critical Chalker-Coddington network model, we need a set of novel techniques that are being developed in a separate paper \cite{SM:DZ-26}. Therefore, part of our derivation here is necessarily heuristic.

To connect with (and build on) standard knowledge, we start from the framework of the nonlinear $\sigma$-model (NL$\sigma$M). The field target space of that model is a coset space $G/K$ with Lie supergroup $G = \mathrm{U}(1,1|2)$ and $K = \mathrm{U}(1|1) \times \mathrm{U}(1|1)$ a compact subgroup of $G$ (this is for the minimal theory with one retarded and one advanced replica, each occurring as a super-multiplet made of one bosonic and one fermionic component). By convention, the coset-valued field $gK$ is realized as a (super-)matrix field $Q$ by conjugating a $K$-invariant saddle-point matrix $Q_0\,$ (a process known as the moment map in symplectic mechanics):
\begin{equation}
    Q(r) = g(r) \, Q_0 \, g(r)^{-1}  = g(r) k(r) \, Q_0 \, k(r)^{-1} g(r)^{-1}  , \quad g(r) \in G , \quad k(r) \in K .
\end{equation}
For present purposes, we make an Iwasawa-style decomposition:
\begin{equation}
    G = N H K,
\end{equation}
where $H = \exp \mathfrak{h}$ is the exponential of a subspace $\mathfrak{h}$ in the orthogonal complement of $\mathrm{Lie}(K)$ in $\mathrm{Lie}(G)$, and $N$ is a nilpotent subgroup normalized by $H$:
\begin{equation}
    h n h^{-1} = n^\prime \quad (h \in H; \; n, n^\prime \in N) .
\end{equation}
More concretely, we adopt the following (non-standard) matrix realization:
\begin{equation}\label{SM:eq:Iwasawa}
    Q_0 = \begin{pmatrix} 0 &\mathbf{1}\cr -\mathbf{1} &0 \end{pmatrix} , \quad h = \begin{pmatrix} A &0\cr 0 &A^{-1} \end{pmatrix} , \quad n = \begin{pmatrix} \mathbf{1} &0\cr B &\mathbf{1} \end{pmatrix} ,
\end{equation}
where $A$ and $B$ are $2 \times 2$ supermatrices with even-type entries $A^0_{\; 0} \in \exp(\mathbb{R})$, $A^1_{\; 1} \in \exp(\mathrm{i}\mathbb{R})$, $B^0_{\; 0} \in \mathbb{R}$, $B^1_{\; 1} \in \mathrm{i}\mathbb{R}$. [It should be noted that the coordinates given by the decomposition $G = N H K$ are not global in the Fermion-Fermion sector. To validate them, one must anticipate symmetry breaking in the infrared limit of the critical theory -- by a mechanism discussed in \cite{SM:Zi-24} for the related case of the $O(3)$ nonlinear $\sigma$-model.] The NL$\sigma$M-field $Q$ is then expressed as
\begin{equation}
    Q = n h Q_0 h^{-1} n^{-1} = \begin{pmatrix}
        -MB &M \cr -M^{-1} - BMB &BM
    \end{pmatrix} , \quad M = A^2 ,
\end{equation}
and the NL$\sigma$M-Lagrangian takes the form
\begin{equation}
    - {\textstyle{\frac{1}{2}}} \mathrm{STr}\, (\nabla Q)^2 =  \mathrm{STr}\, (M^{-1} \nabla M)^2 + \mathrm{STr}\, (M \nabla B)^2 .
\end{equation}
In short, the proposal of \cite{SM:Zi-19} was that $\int d^2r \, \mathrm{STr}\, (M^{-1} \nabla M)^2$ is to be replaced by a Wess-Zumino-Witten (WZW) action $S[M]$ of level $n = 4$ at the renormalization-group fixed point of the critical Chalker-Coddington network model. [To be clear, the WZW anomaly does not arise by renormalization to the infrared; rather, it is already present in the ultraviolet theory, though not obviously so. To detect it, one has to do better than the conventional approach of ``deriving'' (without controlled expansion parameter) the nonlinear $\sigma$-model via Hubbard-Stratonovich transformation, saddle-point approximation, and gradient expansion.]

Because the NL$\sigma$M-Lagrangian is quadratic in $B$, we may integrate over the Gaussian field $B$ to obtain an effective action functional (suitable for $B$-unspecific observables such as the partition function) in terms of $M$ alone. The outcome of this calculation can be anticipated by inspecting the root system of the underlying Lie-algebraic structure. The information relevant for that is encoded in the (right) action of $H$ on $N$:
\begin{equation}
    n \mapsto h^{-1} n h = \begin{pmatrix} A^{-1} &0\cr 0 &A \end{pmatrix} \begin{pmatrix} \mathbf{1} &0\cr B  &\mathbf{1} \end{pmatrix} \begin{pmatrix} A &0\cr 0 &A^{-1} \end{pmatrix}
    = \begin{pmatrix} \mathbf{1} &0 \cr ABA &\mathbf{1} \end{pmatrix} .
\end{equation}
If we take $A$ to be diagonal, $A = \mathrm{diag}(\mathrm{e}^{\varphi/2} , \mathrm{e}^{\mathrm{i}\theta/2})$, then we deduce the transformation law
\begin{equation}
B \mapsto B^\prime  = A B A = \begin{pmatrix}  \mathrm{e}^{\varphi} B^0_{\; 0} &\mathrm{e}^{\frac{1}{2}(\varphi + \mathrm{i}\theta)} B^0_{\; 1} \cr \mathrm{e}^{\frac{1}{2}(\mathrm{i}\theta + \varphi)} B^1_{\; 0} &\mathrm{e}^{\mathrm{i}\theta} B^1_{\; 1}\end{pmatrix} ,
\end{equation}
from which we read off the following system of roots $\alpha$ and (fermion-parity) signed multiplicities $m_\alpha$ for the infinitesimal action:
\begin{equation}
    \alpha \, (m_\alpha) : \quad \varphi \, (+1) , \; \mathrm{i}\theta \, (+1), \; {\textstyle{\frac{1}{2}}} (\varphi + \mathrm{i}\theta) \, (-2) . 
\end{equation}
Two invariants computed from this root system are of special importance:
\begin{equation}
    \sum m_\alpha \alpha = 0 , \quad \sum m_\alpha \alpha^2 = {\textstyle{\frac{1}{2}}} (\varphi - \mathrm{i}\theta)^2 = 2\, \mathrm{STr}^2 (\ln A) .
\end{equation}
The vanishing of the first invariant $(\sum m_\alpha \alpha = 0$) means that the ``radial'' part of the integration measure for $G/K$ is ``flat'' (i.e. independent of $h \in H$) for the decomposition $gK = n h K$. We are highlighting this feature for better contrast with a different situation that will be encountered below.

The information content of the second invariant is the following. It is known \cite{SM:Friedan-85} that the one-loop contribution to the renormalization-group beta function of a NL$\sigma$M with target space $G/K$ is given by the Ricci curvature of $G/K$. Now the invariant sum $\sum m_\alpha \alpha^2 = 2\, \mathrm{STr}^2 (\ln A)$ takes the role of the Ricci curvature in our modified setting with ``slow'' field $h \in H$ and ``fast'' field $n \in N$: by integrating over the fast field $n$, one gets a one-loop correction to the effective action for the slow field $h$: 
\begin{equation}
    S^{1-{\rm loop}} = - \frac{1}{4\pi} \int d^2r \, \sum m_\alpha (\nabla \alpha)^2 = -\frac{1}{2\pi} \int d^2r \, \mathrm{STr}^2 (\nabla \ln A) .
\end{equation}
Assuming the supersymmetric theory to be one-loop exact, one arrives at the fixed-point action $S_\ast [M = A^2]$ proposed in \cite{SM:Zi-19}:
\begin{equation}\label{SM:eq:WZW}
    S_\ast [M] = \frac{n}{8\pi} \int 
    \! d^2r \; \mathrm{STr}\, (M^{-1} \nabla M)^2 + \Gamma_n^{\rm  wzw} - \frac{1}{8\pi} \int \! d^2r \; \mathrm{STr}^2 (\nabla \ln M) \quad (n = 4), 
\end{equation}
with anomalous term $\Gamma_n^{\rm WZW}$ and level $n = 4$. We note that the one-loop correction affects the stiffness constant $n/8\pi$ of the WZW-Witten field $M$. This is important in order for the field theory to match the known phenomenology, as it sets the scaling dimension of $M$ to zero. 

\subsection{Field parameterization underlying the present work}

The present work focuses on the specific observable $\mathbb{E}( |\psi_c(o)|^{2q} )$. Speaking the \emph{lingua franca} of the NL$\sigma$M field integral for the product of one retarded (R) and one advanced (A) Green's function, this observable is expressed as a field-theory expectation value,
\begin{equation}
\mathbb{E}\big( |\psi_c(o)|^{2q} \big) = 2^{-q} \left\langle 
\big( Q_{\rm BB}^{\rm RR}(o) - Q_{\rm BB}^{\rm AA}(o) - \mathrm{e}^{\mathrm{i}\vartheta} Q_{\rm BB}^{\rm RA}(o) + \mathrm{e}^{- \mathrm{i} \vartheta} Q_{\rm BB}^{\rm AR}(o) \big)^q \right\rangle ,
\end{equation}
in the non-compact (or Boson-Boson) sector. The presence of the point contact at position $c$ forces the condition $Q(c) \approx Q_0$ (characteristic of an absorbing boundary), which turns into the exact constraint $Q(c) = Q_0$ in the infrared limit of the critical theory. The phase factor $\mathrm{e}^{\mathrm{i} \vartheta}$ is arbitrary; for definiteness, we set $\vartheta = 0$. 

The field-theory computation simplifies if we choose target-space coordinates that are well adapted to our observable $\mathbb{E}( |\psi_c(o)|^{2q})$. Hence, we modify the choice made in (\ref{SM:eq:Iwasawa}) by reorganizing the Boson-Fermion ordering as
\begin{equation}
\begin{pmatrix}
   {\rm BB} &{\rm BF} &{\rm BB} &{\rm BF} \cr
    {\rm FB} &{\rm FF} &{\rm FB} &{\rm FF} \cr
    {\rm BB} &{\rm BF} &{\rm BB} &{\rm BF} \cr
    {\rm FB} &{\rm FF} &{\rm FB} &{\rm FF}
    \end{pmatrix} \longrightarrow
    \begin{pmatrix}
    {\rm BB} &{\rm BF} &{\rm BF} &{\rm BB} \cr
    {\rm FB} &{\rm FF} &{\rm FF} &{\rm FB} \cr
    {\rm FB} &{\rm FF} &{\rm FF} &{\rm FB} \cr
    {\rm BB} &{\rm BF} &{\rm BF} &{\rm  BB}
    \end{pmatrix} ,
\end{equation}
i.e., we exchange the third and fourth row and column. Notice that $Q_0$ changes as
\begin{equation}
    Q_0 = \begin{pmatrix} 0 &\mathbf{1}\cr -\mathbf{1} &0 \end{pmatrix} \longrightarrow
    \begin{pmatrix} 0 &\sigma_1 \cr -\sigma_1 &0 \end{pmatrix} , \quad \sigma_1 = \begin{pmatrix} 0 &1 \cr 1 &0 \end{pmatrix} .
\end{equation}
We now parameterize the matrix $Q$ in the new basis as follows. Let the central (or radial) factor be
\begin{equation}\label{SM:eq:Q-rad}
    Q_{\rm rad} = \begin{pmatrix}
    0 &0 &0 &\mathrm{e}^\varphi \cr
    0 &0&  \mathrm{e}^{\mathrm{i}\theta} &0 \cr 
    0 &- \mathrm{e}^{-\mathrm{i}\theta} &0 &0 \cr
    - \mathrm{e}^{-\varphi} &0 &0 &0\end{pmatrix} ,
\end{equation}
where $\theta \in \mathbb{R} / 2\pi\mathbb{Z}$, and $\varphi$ is our Gaussian free field \emph{in spe}. The matrix $Q$ is then obtained by conjugating $Q_{\rm rad}$ with the elements $n^\prime$ of a lower triangular (hence nilpotent) group $N^\prime \supset N$,
\begin{equation}\label{SM:eq:nilpot}
    Q = n^\prime Q_{\rm rad}\, n^{\prime -1} , \quad
    n^\prime =  \begin{pmatrix}
    1 &0 &0 &0 \cr
    0 &1 &0  &0 \cr
    \beta &c &1 &0 \cr
    b &\gamma &0 &1
    \end{pmatrix}
    \begin{pmatrix}
    1 &0 &0 &0 \cr
    \eta &1 &0  &0 \cr
    0 &0 &1 &0\cr
    0 &0 &\xi &1
    \end{pmatrix} ,
\end{equation}
where $b, c, \beta, \gamma$ can be viewed as the reshuffled entries of the matrix $B$ parameterizing $N$, and $\xi, \eta$ are a pair of Grassmann variables that have been pulled out of the WZW field $M$. Note that each factor in the product representation of $n^\prime$ is one-step nilpotent, while the group $N^\prime$ is not. [As before, the parameterization of the Fermion-Fermion sector is valid only locally and is justified by symmetry breaking and Cauchy deformation \cite{SM:Zi-24} of the punctured two-sphere.] The standard NL$\sigma$M-Lagrangian (without theta-term) is now expressed as 
\begin{equation}\label{SM:eq:Iwasawa-p}
            - \textstyle{\frac{1}{2}} \mathrm{STr}\, (\nabla Q)^2 = - \textstyle{\frac{1}{2}} \mathrm{STr}\, (\nabla Q_{\rm rad})^2 - \textstyle{\frac{1}{2}} \mathrm{STr}\, \big( [Q_{\rm rad} \, , n^{\prime -1} \nabla n^\prime ] \big)^2 .
\end{equation}
From $- \frac{1}{2} \mathrm{STr} \, (\nabla Q_{\rm rad})^2 = (\nabla \varphi)^2 + (\nabla \theta)^2 $ and the heuristic rule that the $N$-radial part of the NL$\sigma$M-Lagrangian is to be replaced by the WZW-Lagrangian, we expect the (undressed) action functional for $\varphi$ to be 
\begin{equation}\label{SM:eq:S-nodress}
 S_{\rm WZW}[\varphi] = \frac{n}{8\pi} \int_\Sigma \!\! d^2 r \,(\nabla\varphi)^2 .
\end{equation}
We note that the absorbing boundary condition $Q(c) = Q_0$ forces $Q_{\rm rad}(c) = Q_0$ and $n^\prime (c) = \mathbf{1}$. Because $N^\prime$ is lower triangular, the entry in the right upper corner of $Q = n^\prime Q_{\rm rad} \, n^{\prime -1}$ is the entry in the right upper corner of $Q_{\rm rad}$, which is simply $\mathrm{e}^\varphi$. Thus we arrive at the ``bosonization rule''
\begin{equation}\label{SM:eq:bosonize}
\mathbb{E}\big( |\psi_c(o)|^{2q} \big) = 2^{-q} \left\langle 
\big( (Q_{\rm BB}^{\rm RR} - Q_{\rm BB}^{\rm AA} - Q_{\rm BB}^{\rm RA} + Q_{\rm BB}^{\rm AR})(o) \big)^q \right\rangle = \big\langle (\mathrm{e}^{\varphi(o)})^q \big\rangle_{\varphi(c) = 0} \; ,
\end{equation}
which is a major advantage motivating the present $Q$-field parameterization. The subscript in the right-most expression reminds us that the field integral for $\varphi$ is to be computed with the boundary condition $\varphi(c) = 0$.

\subsection{Change of variables}

Since we wish to compute $\langle \mathrm{e}^{q \varphi(o)} \rangle$, we seek a reduction of the full field theory to a reduced one in terms of the field $\varphi$ alone. We propose that such a reduction does exist and is actually tractable! As before, we can guess the outcome of the reduction by inspecting the root system $(\alpha,m_\alpha)$ of the action on $N^\prime$ of the one-parameter group of radial elements $h = {\rm diag}(\mathrm{e}^{\varphi/2}, 1,1, \mathrm{e}^{-\varphi/2})$. This system is
\begin{equation}
    \alpha \, (m_\alpha) : \quad \varphi \, (+1), \; \varphi/2 \, (-4) , 
\end{equation}
where the even-parity root ($\varphi$) is associated with the even variable $b$, and the four odd-parity roots ($\varphi/2$) are associated with the four odd variables $\beta, \gamma, \xi, \eta$. The basic invariants now are
\begin{equation}
    \sum m_\alpha \alpha = - \varphi , \quad \sum m_\alpha \alpha^2 = 0 .
\end{equation}
For reasons given above, the vanishing of the second invariant suggests that the coefficient of the one-loop correction to $\int d^2r \, (\nabla \varphi)^2$ vanishes. Thus, we expect the stiffness constant for the $\varphi$-field to coincide with the WZW-coupling $n/8\pi$ in Eq.~(\ref{SM:eq:S-nodress}), without the correction that occurred in the earlier reduction to $S_\ast[M]$ in Eq.~(\ref{SM:eq:WZW}). 

An important change from before is the non-zero value of the first invariant: the root sum $\sum m_\alpha \alpha = - \varphi$ tells us that the functional integration measure for the radial field $\varphi$  is the formal product $\prod_r \mathrm{e}^{- \varphi(r)} d\varphi(r)$. To eliminate the ill-defined factor $\prod_r \mathrm{e}^{-\varphi(r)}$, we make a change of integration variables
\begin{equation}\label{SM:eq:flatten}
    \xi(r) \to \mathrm{e}^{-\varphi(r)/2} \xi(r) , \quad
     \eta(r) \to \mathrm{e}^{-\varphi(r)/2} \eta(r) .
\end{equation}
The functional ``integration'' measure for the Grassmann fields then transforms as
\begin{equation}
    \prod_r \frac{\partial^2}{\partial \xi(r) \partial \eta(r)} \to \prod_r \mathrm{e}^{+\varphi(r)} \frac{\partial^2}{\partial \xi(r) \partial \eta(r)}
\end{equation}
thus canceling $\prod_r \mathrm{e}^{-\varphi(r)}$. Taking into account the WZW-dynamics of the pair ($\xi, \eta$), the Jacobian of this variable transformation is a (Fujikawa-type) determinant
\begin{equation}
 \Omega[\varphi] / \Omega[0] =
 \Det^\prime \!\left(-\e^{-\varphi/2}\partial_z \, \mathrm{e}^\varphi
 \partial_{\bar z}\, \mathrm{e}^{-\varphi/2}\right) /
  \Det^\prime \!\left(- \partial_z \partial_{\bar z} \right) ,
\end{equation}
where 
\begin{equation}
    z = x + \mathrm{i} y, \quad   \bar{z} = x - \mathrm{i} y, \quad  
     \partial_z = \textstyle{\frac{1}{2}} (\partial_x - \mathrm{i} \partial_y ), \quad    \partial_{\bar{z}} = \textstyle{\frac{1}{2}} (\partial_x + \mathrm{i} \partial_y ) ,
\end{equation}
are complex coordinates and partial derivatives for the torus. The deformed Laplacian here is of the Dolbeault type $\partial_z \partial_{\bar z}$ (instead of the deRham type $\nabla^2$) because the Grassmann fields $\xi$ and $\eta$ (as matrix entries of the WZW-field $M$) are holomorphic resp. anti-holomorphic by the equations of motion for $M$.  The prime on the determinant means that we take it over the $L^2$-space of functions $f$ that obey the Dirichlet condition $f(c) = 0$ [due to $\xi(c) = 0 = \eta(c)$] at the point contact $c$.

Our next steps are to compute $\Omega[\varphi] / \Omega[0]$. For this we abbreviate
\begin{equation}\label{SM:eq:Dolbeault}
    D_\varphi = \mathrm{e}^{\varphi/2}\partial_{\bar z}\circ\e^{-\varphi/2}, \quad
    D_\varphi^\dagger = - \, \mathrm{e}^{-\varphi/2}\partial_z\circ\e^{\varphi/2},
\end{equation}
so that
\begin{equation}\label{SM:eq:Omega} 
\Omega[\varphi] / \Omega[0] = \mathrm{Det}^\prime \left( D_\varphi^\dagger D_\varphi^{\vphantom{\dagger}} \right) / \mathrm{Det}^\prime \left(D_0^\dagger D_0^{\vphantom{\dagger}} \right) .
\end{equation}
A remark to be referenced later is that $\Omega[\varphi]$ is shift-invariant by a constant $\varphi_0\,$:
\begin{equation}
    \Omega[\varphi(\bullet)] = \Omega[\varphi(\bullet) + \varphi_0]
\end{equation}
In other words, $\Omega[\varphi]$ has no dependence on the constant (or zero) mode of $\varphi$.

\subsection{Heat-kernel regularization and variation of determinant}

We begin by writing the log-determinant as a heat-kernel trace with small regularization parameter $\varepsilon$:
\begin{equation}
\ln \frac{\Omega[\varphi]}{\Omega[0]} = \int_\varepsilon^\infty \!\! \frac{dt}{t} \mathrm{Tr}^\prime \big( \mathrm{e}^{-t D_0^\dagger D_0^{\vphantom{\dagger}}} - \mathrm{e}^{-t D_\varphi^\dagger D_\varphi^ {\vphantom{\dagger}}} \big) .
\end{equation}
The prime is to remind us that the trace runs over the $L^2$-space of functions $f$ subject to the Dirichlet condition $f(c) = 0$ at the point contact $c$. Now, for the purpose of deriving a differential equation for $\Omega$, we introduce a real parameter $0 \leq s \leq 1$ by replacing $\varphi(\bullet) \to s \, \varphi(\bullet)$. Then, recalling the definitions (\ref{SM:eq:Dolbeault}) and and using
\begin{equation}
\frac{d}{ds} D_{s \varphi}^\dagger D_{s \varphi}^{ \vphantom{\dagger}} = D_{s \varphi}^\dagger \varphi D_{s \varphi}^{\vphantom{\dagger}} - \frac{\varphi}{2} D_{s \varphi}^\dagger D_{s \varphi}^{\vphantom{\dagger}} - D_{s \varphi}^\dagger D_{s \varphi}^{\vphantom{\dagger}} \frac{\varphi}{2} ,
\end{equation}
we see that the derivative
\begin{equation}\label{SM:eq:diff-eqn}
    \frac{d}{ds} \ln\frac{\Omega[s\varphi]}{\Omega[0]} \equiv A(s) + B(s)
\end{equation}
is a sum of two terms, the first one being
\begin{equation}
A(s) = - \int_\varepsilon^\infty \!\!\!\!\! dt \, \mathrm{Tr}^\prime \varphi D_{s\varphi}^\dagger D_{s\varphi}^ {\vphantom{\dagger}} \mathrm{e}^{-t D_{s\varphi}^\dagger D_{s\varphi}^ {\vphantom{\dagger}}} ,
\end{equation}
and the second one is,
\begin{equation}\label{SM:eq:B-term}
B(s) = \int_\varepsilon^\infty \!\!\!\!\! dt \, \mathrm{Tr}^\prime D_{s\varphi}^\dagger \,\varphi D_{s\varphi}^{\vphantom{\dagger}}\, \mathrm{e}^{-t D_{s\varphi}^\dagger D_{s\varphi}^{\vphantom{\dagger}}} .
\end{equation}
We compute the two terms separately.

\subsection{Computation of the term $A(s)$}

The integrand for $A(s)$ is a total $t$-derivative. So, carrying out the $t$-integral one gets a short-time ($t = \varepsilon$) contribution and another one from late times ($t \to \infty$):
\begin{equation}
    A(s) = \int_\varepsilon^\infty \!\!\!\!\! dt \, \mathrm{Tr}^\prime \varphi \frac{d}{dt} \mathrm{e}^{-t D_{s\varphi}^\dagger D_{s\varphi}^ {\vphantom{\dagger}}} 
    = - \mathrm{Tr}^\prime \varphi \, \mathrm{e}^{-\varepsilon D_{s\varphi}^\dagger D_{s\varphi}^ {\vphantom{\dagger}}}  + \lim_{t \to\infty} \mathrm{Tr}^\prime \varphi \, \mathrm{e}^{- t D_{s\varphi}^\dagger D_{s\varphi}^ {\vphantom{\dagger}}} .
\end{equation}
The latter vanishes because the one-dimensional kernel $\mathbb{C} \cdot \mathrm{e}^{s \varphi/2}$ of $D_{s \varphi}$ is eliminated by the Dirichlet boundary condition $f(c) = 0$, rendering the Laplacian $D_{s \varphi}^\dagger D_{s \varphi}^{\vphantom{\dagger}}$ strictly positive.

The $t=\varepsilon$ term is computed by short-time expansion ($\varepsilon \to 0$) of the heat-kernel diagonal $\langle r |\, \mathrm{e}^{-\varepsilon D_{s\varphi}^\dagger D_{s\varphi}^ {\vphantom{\dagger}}} | r \rangle$. This is a standard procedure, reviewed here for completeness and convenience of the reader. Let $L_s = D_{s\varphi}^\dagger D_{s\varphi}$ for short, and decompose $L_s$ as the sum 
\begin{equation}
 L_s = M_s + \frac{s}{8} \nabla^2 \varphi , \quad \nabla^2 = \partial_x^2 + \partial_y^2 ,
 \end{equation}
of a potential term and a magnetic Laplacian $M_s$:
\begin{equation}
 M_s = -\frac{1}{4} \Big( 
 (\partial_x - \mathrm{i} s a_x)^2  +
 (\partial_y - \mathrm{i} s a_y )^2 
 \Big) , \quad a_x = \partial_y \varphi / 2  , \quad a_y = - \partial_x \varphi / 2 .
\end{equation}
If the condition $f(c) = 0$ were absent, the leading term in the small-$\varepsilon$ asymptotics of $\langle r |\, \mathrm{e}^{-\varepsilon M_s} | r \rangle$ would agree with that of the free Laplacian,
\begin{equation}
    \langle r |\, \mathrm{e}^{\varepsilon \nabla^2 / 4 }| r \rangle = \frac{1}{\pi\varepsilon} ,
\end{equation}
as the presence of a magnetic field $B_{xy} = \partial_x a_y - \partial_y a_x = - s \nabla^2 \varphi /2$ does not change the short-time dynamics in leading order. Now the condition $f(c) = 0$ brings about a change: it pushes the heat-kernel density $d^2r \, \langle r |\, \mathrm{e}^{\varepsilon \nabla^2 / 4 }| c \rangle$ down to zero (by diffusion and absorption) roughly in a disk of radius shrinking as $\sqrt{\varepsilon}$ for $\varepsilon \to 0$. Therefore, the correct small-$\varepsilon$ asymptotics for our situation with Dirichlet condition $f(c) = 0$ is
\begin{equation}
    \langle r |\, \mathrm{e}^{\varepsilon \nabla^2 / 4 }| r \rangle_{f(c) = 0} = \frac{1}{\pi\varepsilon} \Big( 1 - p \big( (r-c)/ \sqrt{\pi\varepsilon} \big) \Big) + \textrm{finite terms} ,
\end{equation}
with some bump function $p(\bullet) \leq 1$. Since the total mass of the bump is not obvious, we treat it as an unknown parameter which is to be determined later (by a consistency argument related to the $\varphi$-zero mode). Noting that 
\begin{equation}
\langle r |\, \mathrm{e}^{-\varepsilon D_{s\varphi}^\dagger D_{s\varphi}^ {\vphantom{\dagger}}} | r \rangle = \langle r |\, \mathrm{e}^{-\varepsilon M_s} | r \rangle \,\mathrm{e}^{- \varepsilon s \nabla^2 \varphi(r) / 8} ,
\end{equation}
we arrive at
\begin{equation}
  - \mathrm{Tr}^\prime \varphi \, \mathrm{e}^{-\varepsilon D_{s\varphi}^\dagger D_{s\varphi}^ {\vphantom{\dagger}}} = - \frac{1}{\pi\varepsilon} \int_\Sigma \!\!   d^2 r \, \varphi(r)  \Big( 1 - p \big( (r-c)/ \sqrt{\pi\varepsilon} \big) \Big) 
\,\mathrm{e}^{- \varepsilon s \nabla^2 \varphi(r) / 8} + \ldots .
\end{equation}
With the total mass of the bump $p$ denoted by $\int d^2r \, p(r-c) = m$, this expression (for $\varepsilon \to 0$) simplifies to
\begin{equation}\label{SM:eq:A-result}
  A(s) = - \mathrm{Tr}^\prime \varphi \, \mathrm{e}^{-\varepsilon D_{s\varphi}^ \dagger D_{s\varphi}^ {\vphantom{\dagger}}} = - \frac{1}{\pi\varepsilon} \int_\Sigma \!\!d^2 r \, \varphi(r) + m \varphi(c) + \frac{s}{8 \pi} \int_\Sigma \!\! d^2 r \,\varphi(r) \nabla^2 \varphi(r) + \mathcal{O}(\varepsilon) .
\end{equation}

\subsection{Computation of $B(s)$ and solution of differential equation}

We turn to the term $B(s)$ of Eq.~(\ref{SM:eq:B-term}). To compute it, we move $D_{s\varphi}^\dagger$ from front to back by cyclicity of the trace, and we then integrate over $t$ to obtain
\begin{equation}
    B(s)  = + \mathrm{Tr}\,  \varphi \, \mathrm{e}^{-\varepsilon D_{s\varphi}^{\vphantom{\dagger}} D_{s\varphi}^ \dagger}  - \lim_{t \to\infty} \mathrm{Tr}\, \varphi \, \mathrm{e}^{- t D_{s\varphi}^{\vphantom{\dagger}} D_{s\varphi}^ \dagger} .
\end{equation}
The subtle point here is that the prime has disappeared ($\mathrm{Tr}^\prime \to
\mathrm{Tr}$), i.e. the Dirichlet condition $f(c) = 0$ is no longer in force. This is because $D_{s \varphi}$ summed with its adjoint $D_{s\varphi}^\dagger$ is to be regarded as a Dirac-type operator mapping between \emph{different} function spaces, and ``trace cyclicity'' actually exchanges the two spaces. [Mathematically speaking, $d\bar{z}\, D_{s \varphi} + {\rm h.c.}$ is a Witten-deformed Dirac operator on the Dolbeault complex of $(0,p)$-forms on $\Sigma$; cf.\ \cite{SM:GKR25}. The Dirichlet condition constrains the $(0,0)$-forms but not the $(0,1)$-forms.)] As a result, the late-time contribution ($t \to \infty$) no longer vanishes and has to be taken into account; this will be done presently.

The computation of the short-time contribution ($t = \varepsilon$)  is almost the same as that for $A(s)$, with the only changes being the sign reversal $\varphi \to - \varphi$ and the absence of the ``absorption hole'' in the heat kernel at $r = c$. Thus, one finds
\begin{equation}
    \mathrm{Tr}\,  \varphi \, \mathrm{e}^{-\varepsilon D_{s\varphi}^{\vphantom{\dagger}} D_{s\varphi}^ \dagger} = \frac{1}{\pi\varepsilon} \int_\Sigma \!\! d^2 r \, \varphi(r)  +
     \frac{s}{8\pi} \int_\Sigma \!\! d^2 r \, \varphi(r) \nabla^2 \varphi(r)
    + \mathcal{O}(\varepsilon) .
\end{equation}
Turning to the long-time contribution to $B(s)$, we observe that the operator $\mathrm{e}^{- t D_{s \varphi}^{\vphantom{\dagger}} D_{s \varphi}^\dagger}$ for $t \to \infty$ becomes a rank-one projector onto the kernel $\mathbb{C} \cdot \mathrm{e}^{-s \varphi/2}$ of $D_{s \varphi}^\dagger$. So,
\begin{equation}
\lim_{t\to\infty} \mathrm{Tr} \, \varphi \, \mathrm{e}^{- t D_{s \varphi}^{\vphantom{\dagger}} D_{s \varphi}^\dagger} = \int_\Sigma \!\! d^2 r \, \varphi(r) \Psi_0(r)^2 ,
\end{equation}
with $\Psi_0(r) \propto \mathrm{e}^{-s\varphi(r)/2}$ the normalized zero mode of $D_{s \varphi}^\dagger$. (We call this zero mode ``virtual'' because it is \underline{not} a zero mode of the operator $D_\varphi^\dagger D_\varphi^{\vphantom{\dagger}}$ whose determinant we are computing.) That limit can be rewritten as a logarithmic derivative:
\begin{equation}
- \lim_{t\to\infty} \mathrm{Tr} \, \varphi \, \mathrm{e}^{- t D_{s \varphi}^{\vphantom{\dagger}} D_{s \varphi}^\dagger} = \frac{d}{ds} \ln \int_\Sigma  \!\! d^2 r \, \mathrm{e}^{-s \varphi(r)} .
\end{equation}
Now, adding up the expressions for $A(s)$ and $B(s)$, we see that the divergent terms ($\sim \varepsilon^{-1}$) cancel to produce a finite result in the limit of $\varepsilon \to 0$:
\begin{equation}\label{SM:eq:AplusB}
    A(s) + B(s) =  \frac{s}{4 \pi} \int_\Sigma \!\! d^2 r \,\varphi(r) \nabla^2 \varphi(r)+  \frac{d}{ds} \ln \int_\Sigma \!\! d^2 r \, \mathrm{e}^{-s \varphi(r) + s m \varphi(c)}  \quad (\varepsilon \to 0) .
\end{equation}
Here we recall that we started the whole calculation by noting the invariance of  $\Omega[\varphi]$ under shifts $\varphi(\bullet) \to \varphi(\bullet) + \varphi_0$ by constants. From Eq.~(\ref{SM:eq:AplusB}) we see that this invariance determines the mass $m = \int d^2 r \, p(r-c)$ of the absorption hole for $A(s)$ to be $m = 1$.

Finally, by integrating the differential equation (\ref{SM:eq:diff-eqn}) from $s = 0$ to $s = 1$, we arrive at
\begin{equation}
     \ln \frac{\Omega[\varphi]}{\Omega[0]} =  \frac{1}{8 \pi} \int_\Sigma \!\! d^2 r \,\varphi(r) \nabla^2 \varphi(r) +  \ln \int_\Sigma \!\! d^2 r \, \mathrm{e}^{-\varphi(r) + \varphi(c)}  - \ln \int_\Sigma \!\! d^2 r,
\end{equation}
or in exponentiated (and partially integrated) form,
\begin{equation}\label{SM:eq:Omega-final}
     \Omega[\varphi] / \Omega[0] =  \mathrm{e}^{-\frac{1}{8 \pi} \int_\Sigma d^2 r \, (\nabla \varphi)^2 } \int_\Sigma \frac{d^2 r}{\mathrm{vol}(\Sigma)} \, \mathrm{e}^{-\varphi(r) + \varphi(c)}  .
\end{equation}
Here it should be stressed that all our calculations were done for the \emph{flat} torus. On a two-dimensional Riemannian manifold with Gauss curvature form $\mathcal{R}$, the short-time expansion of the heat kernel for the Laplacian $\nabla^2 = \mathrm{div} \circ \mathrm{grad}$ (defined by the Riemannian geometry) acquires a correction term:
\begin{equation}
    \langle r |\, \mathrm{e}^{\varepsilon \nabla^2 / 4 }| r \rangle  \, d^2 r = \frac{d^2 r}{\pi\varepsilon} + \frac{\mathcal{R}}{12\pi} + \mathcal{O}(\varepsilon) ,
\end{equation}
which enters the effective action $\ln \Omega[\varphi]$ (also known as the Polyakov anomaly) as a term linear in $\varphi$:
\begin{equation}
    S^\prime = Q \int_\Sigma \! \! \varphi \mathcal{R} ,
\end{equation}
where $Q$ is also referred to as the ``background charge''. Now this form of background charge, which detects curvature, must not be confused with the background charge due to the dipole charge configuration $\int_\Sigma d^2 r \, \mathrm{e}^{-\varphi(r) + \varphi(c)}$ in Eq.~(\ref{SM:eq:Omega-final}), which already exists on a flat surface ($\mathcal{R} = 0$). Truth be told, this distinction was clear neither to the authors of \cite{SM:BWZ-17} nor to the author of \cite{SM:Zi-19}, who overlooked the complication that the zero-mode contributions to $A(s)$ and $B(s)$ need not cancel each other.

\subsection{Reduction to the Gaussian free field}

To complete the process of reduction to a Gaussian free field, we first take care of the constant mode (or zero mode) $\varphi(r) = \varphi_0 = {\rm const}$. By the change of variables (\ref{SM:eq:flatten}) the functional integration measure for the field $\varphi$ has been flattened to $\prod_r d\varphi(r)$. Now we isolate the zero mode $\varphi_0$ and integrate over it with flat measure $d\varphi_0$. Absent in $\ln \Omega[\varphi]$ (and the other parts of the action functional), the zero mode is contained in the observable $\mathrm{e}^{q \varphi(o)}$[cf. Eq.~(\ref{SM:eq:bosonize})] and in the Dirichlet condition $\varphi(c) = 0$ due to absorption at the point contact. Thus the zero-mode integrand to consider is the product $\mathrm{e}^{q \varphi(o)} \delta \big( \varphi(c) \big)$. We shift $\varphi(o) \to \varphi(o) + \varphi_0$ and $\varphi(c) \to \varphi(c) + \varphi_0$ and integrate the product over $\varphi_0$ to obtain
\begin{equation}
    \mathrm{e}^{q \varphi(o)} \int_\mathbb{R} d\varphi_0 \, \mathrm{e}^{q \varphi_0} \delta\big( \varphi(c) + \varphi_0 \big) = \mathrm{e}^{q \big( \varphi(o) - \varphi(c) \big)} .
\end{equation}
Speaking jargon, we see that the elimination of the $\varphi$-zero mode augments the vertex operator $\mathrm{e}^{q\varphi(o)}$ of charge $q$ at the observation point $o$ with a vertex operator $\mathrm{e}^{-q \varphi(c)}$ of the opposite charge $-q$ at the point contact $c$. This charge-neutral configuration is ``dressed'' by the dipole integral (which is charge-neutral, too) from the fermion determinant $\Omega[\varphi]$ in Eq.~(\ref{SM:eq:Omega-final}). So, we expect the conformal scaling limit of the stationary-wave intensity moment  $\mathbb{E}( |\psi_c(o)|^{2q}) \sim  \langle \mathrm{e}^{q \varphi(o)} \rangle_{\varphi(c) = 0}$ to be the three-point function
\begin{equation}\label{SM:eq:3-point}
    \left\langle \mathrm{e}^{q \varphi(o)} \mathrm{e}^{ (1-q) \varphi(c)} \mathrm{e}^{ - \varphi(r)} \right\rangle
\end{equation}
of a Gaussian free field $\varphi$ with zero mode removed [i.e., $\int d^2r \, \varphi(r) = 0$], integrated against $\int_\Sigma d^2r$.  

What is the stiffness constant multiplying $(\nabla \varphi)^2$ in the action functional for the field $\varphi$? It might seem that our initial guess in Eq.~(\ref{SM:eq:S-nodress}) would be renormalized by the factor multiplying the integral on the r.h.s. of Eq.~(\ref{SM:eq:Omega-final}) due to the coupling with the pair $\xi, \eta$. However, this disregards the coupling of $\varphi$ to the other fields ($b, \beta, \gamma$) contained in $n^\prime$; cf. Eq.~(\ref{SM:eq:Iwasawa-p}). By the heuristics of $\sum m_\alpha \alpha^2 = 0$, we expect the latter to produce a counter-term to the one in Eq.~(\ref{SM:eq:Omega-final}), so that the stiffness constant $n/8\pi$ for $\varphi$ actually remains unrenormalized. (Please be assured that any doubt of this scenario is in fact dispelled by our numerical results.) So, the covariance of the Gaussian free field $\varphi$ is
\begin{equation}\label{SM:eq:covar}
 \langle\varphi(r)\varphi(r')\rangle = \frac{4\pi}{n}G(r,r') ,
\end{equation}
with $G$ given explicitly by Eq.~(\ref{SM:eq:Jacobi}) and $\int_\Sigma d^2 r \, G(r,r^\prime) = 0$.

Now the other factor in Eq.~(\ref{SM:eq:Omega-final}) is the integral $\int_\Sigma d^2 r \, \mathrm{e}^{- \varphi(r)}$ due to the virtual zero mode of the Dolbeault pair of WZW-fields $\xi, \eta$. No such infrared contribution is expected from the (non-WZW and non-Dolbeault) fields $(b,\beta,\gamma)$. By the requirement of shift invariance ($\varphi \to \varphi + {\rm const}$), it follows that a modification of the ultraviolet factor $\mathrm{e}^{\varphi(c)}$  cannot arise either. 

In the balance, by the rule of
\begin{equation}
     \left\langle \mathrm{e}^{\sum q_i \varphi(r_i)} \right\rangle = 
     \mathrm{e}^{\frac{1}{2}\sum q_i q_j \big\langle \varphi(r_i) \varphi(r_j) \big\rangle}
\end{equation}
for a Gaussian field $\varphi$, and by taking the cross contractions in the expression (\ref{SM:eq:3-point})  (noting that the self-contractions can be absorbed into the constant normalization factor), we arrive at
\begin{equation}
    M_q(c,o) \propto  \mathrm{e}^{q(1-q) \langle \varphi(c) \varphi(o) \rangle}  \int_\Sigma \!\! d^2 r \; \mathrm{e}^{- q \langle \varphi(o) \varphi(r) \rangle} \mathrm{e}^{ - (1-q) \langle \varphi(c) \varphi(r)\rangle} .
\end{equation}
Finally, by inserting the covariance (\ref{SM:eq:covar}) and including a normalization factor due to the integration over the other fields (not named $\varphi$), we obtain the formula for $M_q(c,o)$ proposed in Eq.~(\ref{SM:eq:torus-M}). 

To finish this subsection, we comment on the behavior under Weyl group transformations ($q \leftrightarrow 1-q)$. As an isometry of the flat torus $\Sigma$, the involutive change of integration variable $r \mapsto c + (o-r)$ leaves the area measure $d^2 r$ invariant. Because it exchanges the difference vectors $r-c$ and $o-r$, it exchanges the two Green's functions $G(r,c) = g(r-c)$ and $G(r,o) = G(o,r) = g(o-r)$. It follows immediately that our expression for $M_q(c,o)$ is Weyl-symmetric. 

\bigskip
\section{Non-parabolicity from omitting the dressing factor}\label{SM:app:fake_nonparabolicity}

The scaling dimension $\Delta_q$ is a quantity predicted by the algebraic structure of a conformal field theory (CFT) for the scaling limit of the IQH transition. Here we distinguish the CFT-given $\Delta_q$ from effective exponents obtained by incomplete fits in finite-size systems.

Approximating the space manifold $\Sigma$ by a lattice, we may write the scaling limit $M_q(c,o)$ as
\begin{equation}\label{SM:eq:ansatz-latt}
\ln M_q^{\rm latt}(c,o) = 4\pi \Delta_q\, g(c-o) + \ln S_q^{\rm latt}(c,o),
\end{equation}
with Green's function $g(r-r^\prime) = G(r,r^\prime)$, defined in Eq.\ (\ref{SM:eq:Poisson}) and shifted by a nonuniversal (but fixed) UV scale $a$. Here $\Delta_q$ is the scaling dimension given in Eq.~\eqref{SM:eq:parabolic}, and the dressing factor $S_q^{\rm latt}$ is a correction present for tori with finite volume. We now argue that $S_q^{\rm latt}$ must be retained for meaningful outcomes in quantitative finite-size fits.

For the slope extraction below, an independent constant log-amplitude is included in the fit. Equivalently, the data and the Green's function may be centered over the fitting window, so that Eq.~(\ref{SM:eq:ansatz-latt}) specifies only the separation-dependent part of the fit. Let
\begin{equation}
g_i=g(c-o_i),
\qquad
s_i=\ln S_q^{\rm latt}(c,o_i),
\end{equation}
for the $N_{\rm fit}$ observation points $o_i$ with $1\leq i \leq N_{\rm fit}$ in the fitting window. The relation may then be written as
\begin{equation}
\ln M_q^{\rm latt}(c,o_i)
=
b_q+4\pi\Delta_q g_i+s_i .
\end{equation}
If it is fitted instead by the incomplete ansatz
\begin{equation}
\ln M_q^{\rm latt}(c,o_i)
\simeq
b_q^{\rm fit}+4\pi\Delta_q^{\rm fit}g_i ,
\end{equation}
with the dressing factor omitted, ordinary unweighted least squares gives
\begin{equation}
\Delta_q^{\mathrm{fit}}-\Delta_q
=
\frac{1}{4\pi}
\frac{\sum_i(g_i-\bar g)(s_i-\bar s)}
     {\sum_i(g_i-\bar g)^2},
\label{SM:eq:slope_projection_SM}
\end{equation}
where
\begin{equation}
\bar g=\frac{1}{N_{\rm fit}}\sum_i g_i,
\qquad
\bar s=\frac{1}{N_{\rm fit}}\sum_i s_i.
\end{equation}
Thus, the bias in $\Delta_q^{\rm fit}$ is the deterministic least-squares projection coefficient of the centered omitted term $s_i-\bar s$ onto the centered fitting direction $g_i-\bar g$. The expression above is for an unweighted fit and is intended to make this projection mechanism transparent. In principle, correlations between data points at different separations can be incorporated through weighted or generalized least squares, with the inverse covariance matrix defining the corresponding fitting inner product. This changes the precise projection coefficient but not the underlying conclusion.

The resulting deviation $\Delta_q^{\rm fit} - \Delta_q$ should not be interpreted as a correction to the scaling dimensions $\Delta_q$ given by CFT; rather, it is simply the finite-window projection of the omitted dressing factor. We note that although $S_q$ is symmetric under $q\to1-q$, $\ln S_q$ is generally not linear in $x_q=q(1-q)$, as $q$ enters inside the exponential before the sum over the background-charge positions is performed. The fit with incomplete ansatz can therefore generate an apparent expansion $\Delta_q^{\rm fit} = a_1x_q+a_2x_q^2+\cdots$, even when the underlying CFT scaling dimensions are exactly parabolic.

\bigskip
\section{Finite-size fits to the analytical prediction}\label{SM:app:fits}

\begin{figure}
\centering
\includegraphics[width=0.85\textwidth]{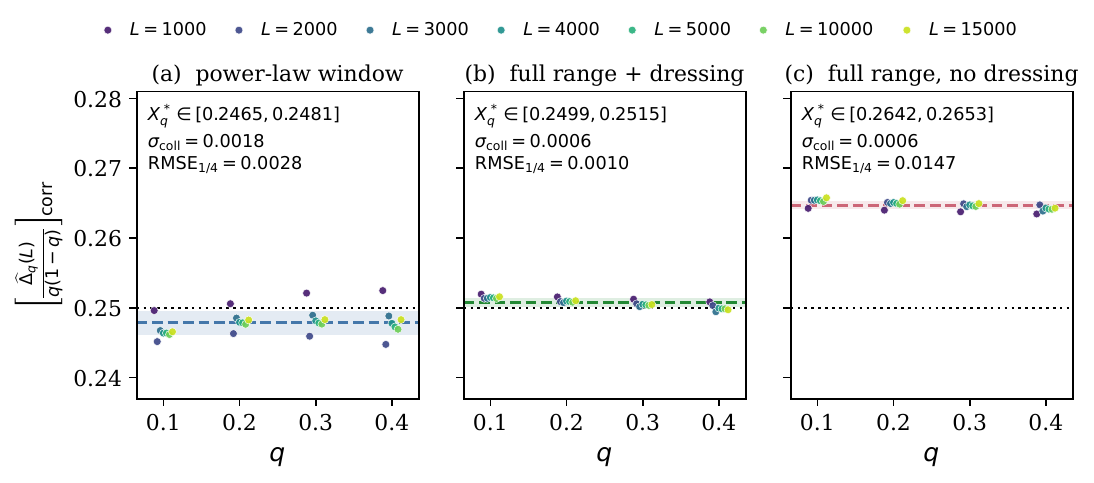}
\caption{Finite-size extraction of the fixed-point coefficient using numerical data for $\mathbb{E}\big( |\psi_c(o)|^{2q} \big)$ averaged over $10^6$ realizations. For the seven displayed sizes, the effective conductivity $\widehat{\sigma}_{xx}(L,q)$ is obtained from (a) a power-law fit over the longest log-binned window satisfying $r\geq4$, $R^2\geq0.9995$, and maximum relative residual $\leq2\%$, with $\widehat{\sigma}_{xx}=q(1-q)/(2\pi\widehat{\Delta}_q)$; or from full-range fits over $2\leq r<L/2$ (b) with and (c) without the dressing factor $S_q$. For each displayed point, its own size is omitted and the remaining six are fitted to $\widehat{\sigma}_{xx}(L,q)=\widehat{\sigma}_{xx}^\ast(q)+b_q/\ln L$. We plot $X_q^{\rm corr}=\{2\pi[\widehat{\sigma}_{xx}(L,q)-b_q^{(-L)}/\ln L]\}^{-1}$. Colors denote $L$, and horizontal offsets are for visibility only. The dotted line marks $1/4$; the colored dashed line and band show the mean and its RMS spread $\sigma_{\rm coll}$. The quoted $X_q^\ast$ interval gives the range of $[2\pi\sigma_{xx}^\ast(q)]^{-1}$ from fits to all seven sizes, and
${\rm RMSE}_{1/4}=\langle(X_q^{\rm corr}-1/4)^2\rangle^{1/2}$.}
\label{SM:fig:fixed_point_test}
\end{figure}

We consider it a major achievement of the present work that we succeeded in fitting the finite-size data in a way that is essentially \emph{parameter-free}. This fit exploited the fortuitous existence of a nearly size-independent crossing point $x_c$ in the distance parameter $x = |c-o|/L$ and the ansatz $\ln F_q(x) -\ln f_{q,L}(x)= \rho(x) / \ln L$ with length-independent $\rho(x)$ and $\rho(x_c) = 0$. Prior to the success with this ansatz, we attempted a finite-size description which is obtained by promoting the fixed-point relation $2\pi\sigma_{xx}^\ast \equiv n=4$, where $\sigma_{xx}^\ast$ is the mean longitudinal conductivity at the renormalization-group fixed point, to a running conductivity $\sigma_{xx}(L)$ in Eq.~(\ref{SM:eq:torus-M}). Lacking theoretical justification, this attempt is recorded here for completeness; it is an empirical scheme to parameterize the finite-size flow and compare the outcomes of different fitting procedures. At the CFT fixed point,
\begin{equation}
    \frac{\Delta_q}{q(1-q)}
    = \frac{1}{2\pi\sigma_{xx}^\ast}
    = \frac14 .
\end{equation}
We characterize the finite-size flow using three fitting procedures. In the first, we extract an effective scaling dimension $\widehat{\Delta}_q(L)$ from the longest logarithmically binned power-law window satisfying $r\geq4$, $R^2\geq0.9995$, and a maximum relative residual of at most $2\%$, and define the corresponding effective conductivity by
\begin{equation}
\widehat{\sigma}_{xx}(L,q)
=
\frac{q(1-q)}
     {2\pi\widehat{\Delta}_q(L)}.
\end{equation}
In the second, we fit the full range $2\leq r<L/2$ directly to the analytical prediction using $\widehat{\sigma}_{xx}(L,q)$ as the fitting parameter. In the third, we repeat the same full-range fit with the dressing factor $S_q$ omitted. For comparison, we parameterize the finite-size dependence obtained from each procedure as
\begin{equation}
\widehat{\sigma}_{xx}(L,q)
=
\widehat{\sigma}_{xx}^\ast(q)
+\frac{b_q}{\ln L}.
\label{SM:eq:sigmaxx_flow}
\end{equation}
Equation~(\ref{SM:eq:sigmaxx_flow}) is used only as an empirical extrapolation of the observed numerical flow. To avoid having each point determine its own finite-size correction, we employ a leave-one-size-out procedure. For a point at size $L$, the remaining six sizes are fitted to Eq.~(\ref{SM:eq:sigmaxx_flow}), yielding $b_q^{(-L)}$, and we plot
\begin{equation}
X_q^{\rm corr}(L)
=
\left\{
2\pi\left[
\widehat\sigma_{xx}(L,q)
-\frac{b_q^{(-L)}}{\ln L}
\right]
\right\}^{-1}.
\label{SM:eq:Xcorr}
\end{equation}
For comparison, we also define the extrapolated fixed-point coefficient
\begin{equation}
X_q^\ast
\equiv
\frac{1}{2\pi{\widehat\sigma_{xx}^\ast(q)}},
\end{equation}
where $\widehat\sigma_{xx}^\ast(q)$ is obtained from a fit of all seven sizes to Eq.~(\ref{SM:eq:sigmaxx_flow}). If the $1/\ln L$ form captures the leading finite-size drift, the corrected values $X_q^{\rm corr}(L)$ should become approximately independent of $L$ and cluster around $[2\pi\sigma_{xx}^\ast(q)]^{-1}=1/4$.

We present the results of the three procedures in Fig.~\ref{SM:fig:fixed_point_test}. The power-law extraction in panel (a) gives a reasonably narrow collapse, but with a larger spread and a small downward shift from $1/4$, yielding $X_q^\ast\in[0.2465,0.2481]$, $\sigma_{\rm coll}=0.0018$, and ${\rm RMSE}_{1/4}=0.0028$. The full-range fit to the analytical prediction in panel (b) gives a substantially tighter collapse and an extrapolated interval $X_q^\ast\in[0.2499,0.2515]$ containing $1/4$, with $\sigma_{\rm coll}=0.0006$ and ${\rm RMSE}_{1/4}=0.0010$. By contrast, omitting $S_q$ in panel (c) yields an equally narrow collapse, $\sigma_{\rm coll}=0.0006$, but around $X_q^\ast\in[0.2642,0.2653]$, well separated from $1/4$, with ${\rm RMSE}_{1/4}=0.0147$. This comparison should not be taken to imply that all finite-size deviations originate from the dressing factor. Omitting $S_q$ introduces a specific geometric bias, while separate corrections remain from the logarithmically slow approach to the scaling limit. The full-range fit including $S_q$ controls the former and extrapolates to a fixed-point coefficient consistent with $1/4$, but in view of the unexplained $1/\ln L$ behavior we cannot be satisfied with it as a consistent description of the renormalization-group flow.

\end{document}